\RequirePackage[hyphens]{url}
\documentclass[mnsc,nonblindrev]{informs3h}
\OneAndAHalfSpacedXI

\usepackage[round]{natbib}
\usepackage{multibib}
\def\bibfont{\small}

\newcites{App}{References for Appendix}

\usepackage{secdot}
\usepackage{epsfig}
\usepackage{amssymb}
\usepackage{amsmath}
\usepackage{color}
\usepackage{graphicx}
\usepackage{multirow}
\usepackage{enumerate}
\usepackage{subfig}
\usepackage{natbib}

 \bibpunct[, ]{(}{)}{,}{a}{}{,}%
 \def\bibfont{\small}%
\usepackage[ruled,linesnumbered]{algorithm2e}
\usepackage[normalem]{ulem}
\usepackage{float}
\usepackage{enumitem}
\usepackage{amsmath}
\usepackage{epstopdf}
\usepackage{graphicx}
\usepackage{tikz}
\usepackage{pgfplots}
\usepackage{subfig}
\pgfplotsset{compat=newest}
\usepackage{pgfplotstable}
\usetikzlibrary{arrows.meta,babel,calc,intersections,backgrounds,patterns,plotmarks,decorations.pathreplacing}
\usepackage{multirow}
\usepgfplotslibrary{fillbetween}
\tikzstyle{every picture} += [>=stealth]
\makeatother
\usepackage{lipsum}
\usepackage[justification=centering]{caption}
\usepackage{booktabs}
\usepackage[hyphens]{url}
\usepackage{mathtools}

\TheoremsNumberedThrough    
\ECRepeatTheorems
\EquationsNumberedThrough  
\makeatletter
\newcommand\footnoteref[1]{\protected@xdef\@thefnmark{\ref{#1}}\@footnotemark}
\makeatother
\usepackage{apptools}
\usepackage{bbm}
\usepackage{chngcntr}
\usepackage{lipsum}
\usepackage{float}
\usepackage{hyperref}
\usepackage{bm}
\hypersetup{
    hidelinks,
    colorlinks=true,
    linkcolor=black,
    citecolor=blue
} 

\usepackage{enotez}
\let\footnote=\endnote
\setenotez{backref = true}
\setenotez{list-name = {Endnotes}}
\usepackage[hang,flushmargin]{footmisc}
\makeatletter
\def\footnoterule{\relax
  \kern 1pt
  \hbox to \columnwidth{\vrule width 0.5\columnwidth height 0.5pt\hfill}
  \kern 1pt}
\makeatother

\tikzset{
 hatch distance/.store in=\hatchdistance,
 hatch distance=10pt,
 hatch thickness/.store in=\hatchthickness,
 hatch thickness=2pt
 }
 \makeatletter
 \pgfdeclarepatternformonly[\hatchdistance,\hatchthickness]{flexible hatch}
 {\pgfqpoint{0pt}{0pt}}
 {\pgfqpoint{\hatchdistance}{\hatchdistance}}
 {\pgfpoint{\hatchdistance-1pt}{\hatchdistance-1pt}}%
 {
 \pgfsetcolor{\tikz@pattern@color}
 \pgfsetlinewidth{\hatchthickness}
 \pgfpathmoveto{\pgfqpoint{0pt}{0pt}}
 \pgfpathlineto{\pgfqpoint{\hatchdistance}{\hatchdistance}}
 \pgfusepath{stroke}
 }
\makeatother
\usetikzlibrary{shapes.geometric, arrows, calc, shapes.misc, matrix, shapes, arrows, calc, positioning, shapes.geometric, shapes.symbols, shapes.misc, intersections, chains, scopes}
\tikzstyle{node1} = [circle, circle sides=6,minimum width=1.25cm, text badly centered, text width=2.35em,minimum height=1.15cm, draw=black, font=\small]
\tikzstyle{node2} = [rectangle,rounded corners, minimum width=1.25cm, minimum height=1cm, text centered, draw=black, font=\small]
\tikzstyle{arrow} = [thick,->,>=stealth]

\begin{document}

\RUNAUTHOR{Li et al.}
\RUNTITLE{Privacy Preserving Adaptive Experiment Design}
\TITLE{Privacy Preserving Adaptive Experiment Design\footnote{Authors are listed in alphabetical order.}}
\ARTICLEAUTHORS{
 \AUTHOR{Jiachun Li}
 \AFF{Laboratory for Information and Decision Systems, MIT, \url{jiach334@mit.edu}}
 \AUTHOR{Kaining Shi}
 \AFF{Zhili College, Tsinghua University, \url{skn20@mails.tsinghua.edu.cn}}
 \AUTHOR{David Simchi-Levi}
 \AFF{Laboratory for Information and Decision Systems, MIT, \url{dslevi@mit.edu}}
}

\ABSTRACT{%
\textbf{Abstract.}
Adaptive experiment is widely adopted to estimate conditional average treatment effect (CATE) in clinical trials and many other scenarios. While the primary goal in experiment is to maximize estimation accuracy, due to the imperative of social welfare, it's also crucial to provide treatment with superior outcomes to patients, which is measured by regret in contextual bandit framework. These two objectives often lead to contrast optimal allocation mechanism.  Furthermore,  privacy concerns arise in clinical scenarios containing sensitive data like patients health records. Therefore, it's essential for the treatment allocation mechanism to incorporate robust privacy protection measures. In this paper, we investigate the tradeoff between loss of social welfare and statistical power in contextual bandit experiment. We propose a matched upper and lower bound for the multi-objective optimization problem, and then adopt the concept of Pareto optimality to mathematically characterize the optimality condition. Furthermore, we propose \textit{differentially private} algorithms which still matches the lower bound, showing that privacy is "almost free". Additionally, we derive the asymptotic normality of the estimator, which is essential in statistical inference and hypothesis testing.

}
\maketitle
\vspace{-2mm}
\section{Introduction}\label{sec: intro}

\subsection{Background}\label{subsec:background}

The contextual bandit framework, a prominent and effective approach for sequential decision-making, is distinguished by its adaptability in progressively refining decisions based on accumulating information. This stands in contrast to reliance on static, offline datasets and batch learning methodologies. While most literature focus on developing algorithms like UCB or Thompson sampling to minimize the cumulative loss of rewards (like revenue or social welfare), it has been shown in \cite{lai1985asymptotically} that 
adaptive allocation strategies can surpass the efficiency of certain conventional random experimental approaches, such as Randomized Control Trials (RCTs)and has drawn much attention in recent experiment design works (\citealt{zhao2023adaptive}, \citealt{dai2023clip}).

Considering the following motivating example of clinical trials.
These trials necessitate evaluating the efficacy of new pharmaceutical interventions across diverse patient circumstances. 
Regret here is measured as the
cumulative detriment to patient welfare, thus necessitating its minimization. This imperative becomes particularly acute in the case of rare or fatal diseases, where the goal is to administer the most effective treatment possible to patients within the trial.
The heterogeneity of patient profiles, characterized by diverse attributes such as age, gender, and genotype, significantly influences the efficacy of treatment. 
Therefore, it's crucial to evaluate the efficacy of drugs across varied patient profiles to identify treatments with superior therapeutic benefits while mitigating potential adverse effects for specific patient groups.
  This illustrates the necessity to estimate the conditional average treatment effect (CATE) (see \citealt{abrevaya2015estimating}, \citealt{fan2022estimation}, \citealt{wager2018estimation}) in adaptive allocation assignment problems while keeping the loss of welfare, or regret at minimum. This dual focus on minimizing regret and accurately estimating CATE is central to both experimental design and contextual bandits in academic literature.

While online regret minimization and statistical inference have been extensively studied separately, the simultaneous pursuit of these objectives introduces novel complexities. This duality of purpose can result in conflicting optimal allocation strategies, as illustrated in some recent works (\citealt{simchi2023multi}). In specific, better accuracy of statistical inference 
typically necessitates broader exploration of various treatment options. This exploration entails more frequent engagement with suboptimal treatment arms, resulting in increased regret.
 In converse, a focus on minimizing regret restricts the algorithm’s engagement with suboptimal arms, which in turn limits the scope of exploration necessary for robust statistical inference.   Moreover, the presence of patient-specific features and heterogeneous treatment effects introduce further complexity.
The estimation and inference for one subgroup of patients cannot be transferred to another, and the arrival of various types of patients may be highly non-stationary, which complicates the inference process for certain subgroups due to insufficient data. While the tradeoff of regret and estimation accuracy has been clearly characterized in homogeneous ATE setting, it remains unknown for the conditional average treatment case when covariates are present. We formulate it in the following question:

\textit{
\textbf{Question 1:} Given a budget of welfare loss (or regret), what's the best possible accuracy of estimation for CATE and how can we achieve such an accuracy?
}

Privacy concerns arise in scenarios involving sensitive data types, such as healthcare records, financial information, or digital footprints.  Utilizing algorithms to mine population-wide patterns without incorporating privacy safeguards can inadvertently expose individual private details(\citealt{carlini2019secret}, \citealt{melis2019exploiting}, \citealt{niu2022differentially}). Such privacy risks also extend to the estimation of Conditional Average Treatment Effect (CATE), as the flexibly estimated CATE function has the potential to inadvertently disclose sensitive individual information, including covariates, treatment statuses or outcomes.
Differential Privacy (DP) has been established as a rigorous mathematical framework for defining privacy, which has gained widespread adoption in practices by major organizations such as the U.S. Census Bureau and companies like Apple and Google for data publication and analysis (\citealt{erlingsson2014rappor}, \citealt{abowd2018us}). DP provides a robust protection against privacy attacks, even when an
adversary possesses substantial external information. (\citealt{dwork2006calibrating})
 However, it is widely known that in the realms of both statistical estimation and regret minimization, "\textit{Privacy comes at a cost}".
While it is well established for DP-statistical estimation with offline, identically and independently distributed data, it's much more subtle to perform valid DP-estimation for online, potentially correlated data. Similarly, it has been a long standing problem to develop a differentially private algorithm for regret minimization in contextual bandit framework.
Considering our objective to design an allocation mechanism that achieves enhanced accuracy while simultaneously minimizing regret,a DP version of our mechanism necessitates a delicate balance of these dual tasks. This leads to a critical inquiry: to what extent must one incur a 'cost' to ensure privacy while striving to optimize both accuracy and regret minimization?

\textit{
\textbf{Question 2:} With the constraint that the experimenter need to protect the privacy of participants, is it still possible to attain the same estimation accuracy as well as social welfare loss?
}

To the best of our knowledge, our work is the first one to handle these two tasks simultaneously in a DP manner.

\subsection{Problem Formulation}
\ \ \
In adaptive experiment design with heterogeneous treatment effect, there is a binary set $\mathcal{A}=\{0,1\}$ of arms (i.e., treatments or controls) and a finite feature set $\mathcal{X}=\{X_1,X_2,\cdots,X_M\}$ with $|\mathcal{X}|=M$.  Suppose $n$ is the time horizon (or the total number of experimental units). At each time $t \leq n$, for every arm $a \in \mathcal{A}$ and feature of the unit $x \in \mathcal{X}$, we can observe a reward (outcome) $r_t(a|x)$. We assume that the features follow a sequence of discrete distributions $P_X=\{P_X^t\}_{t\geq1}$, where $P_X^t=(p_1^t,\cdots, p_M^t)\in (0,1)^M (\Sigma_{j=1}^M p_j^t=1\ \forall t\geq 1)$, which means the probability of observing experimental unit with feature $X_j$ at time $t$ is $p_j^t$. Denote $f_j(m):=\mathbb{E}\left[\sum_{1\leq t\leq m}1_{\{x_t=X_j\}}\right]=\sum_{1\leq t\leq m}p_j^t$, which represents the expected number of occurrences of feature $X_j$ in the first $m$ periods, for any $1\leq j\leq M$ and $1\leq m\leq n$. We have the following assumption for the distribution of features.

\begin{assumption} \label{assumption}
    \textbf{Seasonal Nonstationarity}\\
(1)$\exists C_1, C_2>1$, s.t. $\forall 1 \leq j\leq M$, $C_1<\frac{f_j(n)}{f_j(\frac{n}{2})}<C_2$.\\
(2)$f_{min}(n):=\min_{1\leq j\leq M} f_j(n)\geq \Omega(\log n)$
\end{assumption}

Intuitively, this assumption says in the first and the second half periods, every features will be expected to appear approximately same and at least $\Omega(\log n)$ times. Compared to the simplest assumption where the distribution of features for patients is stationary at each time period, our assumption greatly expands the applicability of our method. For example, different types of patients may have completely different patterns of occurrence on weekdays and weekends, or during different seasons. In this case, the simple assumption that patients' features have a stable distribution will no longer hold, but this situation may still conform to our non-stationary seasonal assumption.

\begin{remark}
   Though here we denote it as non-stationarity, in fact our assumption is very mild and contains oblivious adversarial case. To see this, note that we allow the distribution $P_t^X$ to be arbitrary, so for any oblivious adversarial sequence $(H_t)_{t=1}^n$ satisfying assumption 1, we can just set 
   $p_i^t=1$ for $H_t=X_i$ and $p_j^t=0$ for any $j \neq i$ to obtain the desired sequence.
\end{remark}


After observing feature $x_t\in \mathcal{X}$ and choosing arm $a_t\in \{0,1\}$ based on a policy $\pi$, the reward of the chosen arm $r_t:=r_t\left(a_t|X_t\right) \in[0,1]$ can be observed. A stochastic MAB instance can be denoted by $\nu=\left(P_X, P_0(X_1),P_1(X_1) \cdots, P_0(X_M),P_1(X_M)\right)$, where $P_i(X_j)$ is the distribution of the rewards of arm $i$ and feature $X_j$ and $P_X$ is the distribution of features. We define the conditional average treatment effect (CATE) of a feature $x$ as $\Delta(X_j):=\mathbb{E}\left[r_t(1|X_j)\right]-\mathbb{E}\left[r_t(0|X_j)\right]$, for any $X_j \in \mathcal{X}$. We also denote $\sigma_{ji}=\mathbb{V}\left[r(i| X_j)\right]$ for $i=0,1$ and $j=1, \cdots, M$ as the variance of reward of playing arm $i$ when facing context $X_j$. In this paper, we will elaborate on $\left|\Delta(x)\right|=\Theta(1)$ for all $x \in \mathcal{X}$, which is arguably the most fundamental case in real applications. Denote all stochastic MAB instances satisfying the mentioned assumptions to constitute a feasible set $\mathcal{E}_0$.

A key index to measure the efficiency of online learning policy $\pi$ is accumulative regret, defined as the expected difference between the reward under the optimal policy and the policy $\pi$, i.e., $\mathcal{R}(n, \pi)=\mathbb{E}^\pi\left[\sum_{i=1}^n \left[r_i(a^*(X_i)|X_i)- r_i\left(a_i|X_i\right)\right]\right]$, where $a^*(X_i)$ is the optimal arm of feature $X_i$. In addition, an admissible adaptive estimator $\hat{\Delta}(X_j)$ maps the history $\mathcal{H}_n$ to an estimation of $\Delta(X_j)$. We use the error defined as the mean square error of $\Delta(X_j)$ and $\hat{\Delta}(X_j)$, i.e., $\left.e\left(n, \hat{\Delta}(X_j)\right)=\mathbb{E}\left[\left(\Delta(X_j)-\hat{\Delta}(X_j)\right)^2\right]\right)$ to measure the quality of the estimation. We define $\hat{\Delta}:=\{\hat{\Delta}(X_j)\}_{1\leq j \leq M}$ to represent all the estimators on the gap between two arms for each feature. A design of an MAB experiment can then be represented by an admissible pair $(\pi, \hat{\Delta})$.
Different from traditional MAB problems, the optimal design of MAB experiments in this paper is solving the following minimax multi-objective optimization problem:
\begin{equation}
\min _{(\pi, \hat{\Delta})} \max _{\nu \in \mathcal{E}_0}\left(\mathcal{R}_\nu(n, \pi), \max _{1\leq j \leq M} e_\nu\left(n, \hat{\Delta}(X_j)\right)\right)
\end{equation}
where we use the subscript $\nu$ to denote the MAB instance. Eq. (1) mathematically describes the two goals: minimizing the regret and the largest estimation error among all features. 


The above is a rigorous mathematical description of our first question that we previously presented. 
This sets the stage for our second question, which concerns about the price of protecting privacy for both regret and CATE estimation, and how it will affect the balance between minimizing regret and estimation error. 
In order to rigorously address this question, we first need the following definition of differential privacy, which was first proposed by \cite{shariff2018differentially} and then widely adopted in DP-contextual bandit problems:

\begin{definition}
(\textbf{$(\varepsilon, \delta)$-anticipating private contextual bandit algorithm}).\\
A bandit algorithm $\pi$ is $(\varepsilon, \delta)$-private if for two neighboring datasets $D=\{(x_i,a_i,r_i)_{i=1}^n\}$ and $D^{\prime}=\{(\hat{x}_i,\hat{a}_i,\hat{r}_i)_{i=1}^n\}$ of feature, action and reward pairs that differs in at most one time step $t$, then for all $S \subseteq \mathcal{A}^{n-t}$:
$$
\pi\left(a_{t+1}, \cdots, a_n \in S \mid D\right) \leq e^\varepsilon\pi\left(a_{t+1}, \cdots, a_n \in S \mid D^{\prime }\right) +\delta \nonumber
$$
where $\mathcal{A}=\{0,1\}$ is the set of actions.
\end{definition}

This definition is slightly different with the classical differential privacy (DP). \cite{shariff2018differentially} propose a notion of "joint DP" in the context of linear contextual bandits and is later adopted by \cite{chen2022privacy} as anticipating DP (ADP). The key difference of ADP is to restrict the output sets as allocations strictly after a patient of interest at time $t$. Such a restriction is motivated by two reasons. The first one is that following the classical DP will inevitably lead to linear regret. The second reason is that classical DP assumes that the adversary has access to the provided action $a_t$ at time $t$, which is unreasonable in most adaptive experiments, as communication about $\left(x_t, a_t, r_t\right)$ at time $t$ is expected to be secured and the data prior to time $t$ have no impact on the privacy of patient $t$ because the decision making algorithm has no knowledge of $x_t$ before time $t$. Therefore, only the privacy of outputs \textit{after} time $t$ needs to be protected. For a more detailed discussion about ADP, one can refer to \cite{chen2022privacy}.





\noindent{\textbf{Technical Difficulties and Our Contribution}}

\noindent \textbf{1. Learning and Balancing the Length of RCTs for Different Feature}
As claimed in \cite{simchi2023multi}, the key idea of balancing regret and estimation error is to properly set the length of random  control trials (RCT). However, in our setting each feature may vary enormously in their occurrences, and it can also be highly non-stationary. Since we are interested in the worst estimation among all features, we should set the length of RCTs for all features to be the same as that of the feature with least occurrence frequency, i.e, $f_{min}(n)$. Since we don't know $f_{min}(n)$ at the beginning of experiment, by the assumption of seasonal non-stationarity, we propose an algorithm named ConSE, which divides the experiment into two phase: in the first half periods, it chooses to minimize regret while learning the frequency of occurrences $f_j(\frac{n}{2})$, and in the second half periods ConSE runs RCT for $\hat{f}_{min}(\frac{n}{2})$ periods for each feature, which is estimated from the first phase. 

Another contribution of our result is the improvement of analysis compared to existing work(\citealt{simchi2023multi}). In particular, we develop a tighter upper bound which is tight up to constant, which helps to have a better characterization of the Pareto optimal curve for regret and estimation error. Besides, the proposed estimator in this paper is asymptotically normal, which is vital in constructing confidence interval and testing hypothesis and has been a long standing issue for adaptive experiment design literature(\citealt{simchi2023multi}, \citealt{dai2023clip}, \citealt{zhao2023adaptive}). See section 3 for a more detailed discussion.

\noindent \textbf{2. Privatizing Feature Information in Non-stationary Environment }

Differential privacy is known to be more difficult in bandit setting due to its highly correlated actions. For multi-arm bandit, algorithms based on tree mechanism proposed in \cite{chan2011private} have been proved to be optimal up to polylog factors(see \citealt{tossou2016algorithms}, \citealt{azize2022privacy}, \citealt{sajed2019optimal}). However, when it comes to contextual bandit, things become more complicated, as the algorithm not only needs to privatize the reward of each arm, but also the context of each patient.
Most existing works focus on setting where reward function is in a specific function class like (generalized) linear function (\citealt{hanna2022differentially}, \citealt{shariff2018differentially}, \citealt{zheng2020locally}, \citealt{chen2022privacy}).
However, in clinical trials, it's risky to believe the treatment effect of one type of customer can be generalized to other types in certain way (like a linear function). Therefore, in this paper we don't assume any structure of CATE among different types of patients, which forces us to propose mechanisms different from existing literature. The second difficulty arises from the non-stationarity assumption, which has been considered in very few works. In particular, this rules out the possibility of merging different features as a whole and applying a unified mechanism to them. 

To conquer all the difficulty mentioned, we propose a "Double Privacy" algorithm, which treats each feature separately and 
doubly privatize the patients' information:
first of all, we use the idea of tree mechanism and divide the whole experiment into batches, where the estimation of rewards are only updated at the end of each batches.
Secondly, we randomize the length of each batch to protect the context information, which is, to our best knowledge, novel in DP-contextual bandit setting.
Finally, our "Double Privacy" allows experimenters to balancing regret and the estimation accuracy of CATE in any given level, and our subsequent theoretical guarantees ensure that no method can simultaneously outperform our algorithm in minimizing regret and accurately estimating CATE.

\subsection{Literature Review}
\noindent\textbf{Adaptive Experiment Design}

Experimental design is
 witnessing a surge in popularity across operations research,
econometrics, and statistics (see, e.g., \citealt{johari2015always}, \citealt{bojinov2021panel}, \citealt{bojinov2023design},\citealt{xiong2023optimal})
Adaptive experimental design emerges as a particularly relevant area to our current focus(\citealt{hahn2011adaptive}, \citealt{atan2019sequential}, \citealt{greenhill2020bayesian}). MAB itself can also be viewed as
adaptive experimental design, but with the objective of minimizing regret, while most adaptive experiment design literature primarily concentrates on (conditional) ATE.
\cite{kato2020efficient} investigate adaptive experiments for ATE when covariates are observable. \cite{qin2022adaptivity}
delve into bandit experiments influenced by a potentially nonstationary sequence of contexts, proposing a unified estimator robust against the contextual impact. 
There are some recent works trying to demonstrate the statistical superiority of adaptive experiment compared classical non-adaptive experiment design, where the measurement of precision is the (asymptotic) variance of the estimator. 
In \cite{dai2023clip}, they propose a measurement called Neyman regret, and show that an adaptive design with asymptotically optimal variance is equivalent to sub-linear Neyman regret, thus transforming it into a regret minimization problem.
\cite{zhao2023adaptive} consider a similar setting, but adopt a competitive analysis framework.

Another emerging field is multitasking bandit problems, where minimizing regret is not the only objective (see, e.g., \citealt{yang2017framework}, 
\citealt{yao2021power}, \citealt{zhong2021achieving}). 
\cite{erraqabi2017trading} also consider the tradeoff between regret and estimation error, and propose a new loss function combining these two objectives together. The most related work to this paper may be \cite{simchi2023multi}, where they consider the tradeoff between regret and ATE estimation. We extend their framework to contextual bandit setting, derive a similar Pareto optimality characterization, and consider the additional requirement to protect patients' privacy.

\noindent \textbf{Differentially Private (Contextual) Bandit Learning}

Differential privacy (\citealt{dwork2006calibrating}) has emerged as gold-standard
for privacy preserving data-analysis, as it ensures that the output of the data-analysis algorithm has minimum dependency on any individual datum. Differentially private variants of online learning algorithms have been successfully developed in various settings (\citealt{guha2013nearly}), including a private UCB-algorithm for the MAB problem ( \citealt{azize2022privacy}, \citealt{tossou2016algorithms}) as well as
UCB variations in the linear bandit settings (\citealt{hanna2022differentially}, \citealt{shariff2018differentially}) 
The UCB algorithm maintains a dynamic,
high-probability upper-bound for each arm’s mean, and at each timestep optimistically
pulls the arm with the highest bound. The above-mentioned $\varepsilon$-differentially private ($\varepsilon$-DP)
analogues of the UCB-algorithm follow the same procedure except for maintaining noisy
estimations using the “tree-based mechanism” (\citealt{chan2011private}, \citealt{dwork2010differential}).
This mechanism functions by continuously releasing aggregated statistics over a stream of T observations,
introducing only $\frac{\text{polylog}(T)}{\varepsilon}$ noise in each timestep, and thus leading to an added pseudo regret of order $\frac{\text{polylog}(T)}{\varepsilon}$.
It was shown in \cite{shariff2018differentially} that any $\epsilon$-DP stochastic MAB algorithm must incur an added pseudo regret of $\Omega(\frac{K\log(T)}{\epsilon})$, and this lower bound is matched by \cite{sajed2019optimal}, using a  batched elimination algorithm. 

However, when it comes to DP-contextual bandit, 
so far there isn't a golden standard that works for general contextual bandit problems. Instead, most works focus on contextual linear bandit (\citealt{shariff2018differentially}, \citealt{hanna2022differentially},\citealt{charisopoulos2023robust}) and adopt a relaxed definition named \textit{joint-DP} or \textit{anticipating-DP}. 
These works are in general variants of Lin-UCB(\citealt{abbasi2011improved}), which is known to be optimal for contextual linear bandits. Thanks to the well-known post-processing theorem in differential privacy, they can make use of the special structure of linearity and turn to privatize the matrix of contexts and actions. A lower bound for contextual linear bandit was proposed in \cite{shariff2018differentially}, and then was matched in \cite{shariff2018differentially}, \cite{hanna2022differentially} up to polylog factors. \cite{chen2022privacy} consider differential privacy in dynamic pricing problem in a generalized linear model. They privatize the covariance matrix and use MLE to estimate the model parameters. A follow-up work (\citealt{chen2021differential}) considers dynamic pricing in a non-parametric model and derive an upper bound of $\tilde{O}\left(T^{(d+2) /(d+4)}+\varepsilon^{-1} T^{d /(d+4)}\right)$, which is not known to be optimal or not. There are also other works focusing on other definition of differential privacy, like local-DP (\citealt{han2021generalized}, \citealt{zheng2020locally}, \citealt{ren2020multi}) or shuffle-DP (\citealt{tenenbaum2021differentially}, \citealt{hanna2022differentially},\citealt{chowdhury2022shuffle}) model.

\noindent \textbf{Differentially Private Estimation and Inference}

There has been some initial work on differentially private causal inference methods. \cite{lee2013neighborhood} proposed a privacy-preserving inverse propensity score estimator for estimating average
treatment effect (ATE). \cite{komarova2020identification} study the ramifications of differential privacy on the identification of statistical models
and demonstrate the obstacles encountered in regression discontinuity design with
privacy constraints. In \cite{kusner2016private}, the authors focus on privatization of statistical dependence scores (such as Spearman’s $\rho$ and Kendall’s $\tau$). The main focus of  is to derive privatised
scores that still can accurately  discern the causal direction between two random variables. 
\cite{agarwal2021causal} study parametric estimation of causal
parameters under the local differential privacy framework.

When it comes to adaptive experiment, to our knowledge there is no similar work trying to estimating CATE privately. In the meanwhile, there are some works trying to demonstrate the impact of differential privacy in adaptive data analysis. \cite{nie2018adaptively} prove
that most bandit algorithms including UCB and Thompson Sampling lead to negative bias in empirical means. Interestingly, in \cite{neel2018mitigating} it is shown that the data gathered in DP-manner can help to mitigate bias of estimation.
For a more detailed summary of DP statistical inference, one can refer to \cite{kamath2020primer}.

\section{A Warm-up: Upper and Lower Bound Without Privacy Constraint}\label{sec:warmup}

In this section, we aim to answer the first question proposed in subsection \ref{subsec:background}, i.e. \textit{\textbf{what's the best possible accuracy of estimation for CATE given a budget of regret}}, by first showing a lower bound and then proposing an algorithm \textbf{ConSE} with matching upper bound. Besides, we also use this section as a warm-up to describe the technical difficulties of this problem and how to conquer them, which can be helpful to understand the more complicated algorithm in section 3 with privacy constraints.
In the following theorem,
we provide a mini-max lower bound to 
explicitly show the best possible estimation accuracy with a constraint on regret budget.

\begin{figure}
    \centering
    \includegraphics[width=0.5\linewidth]{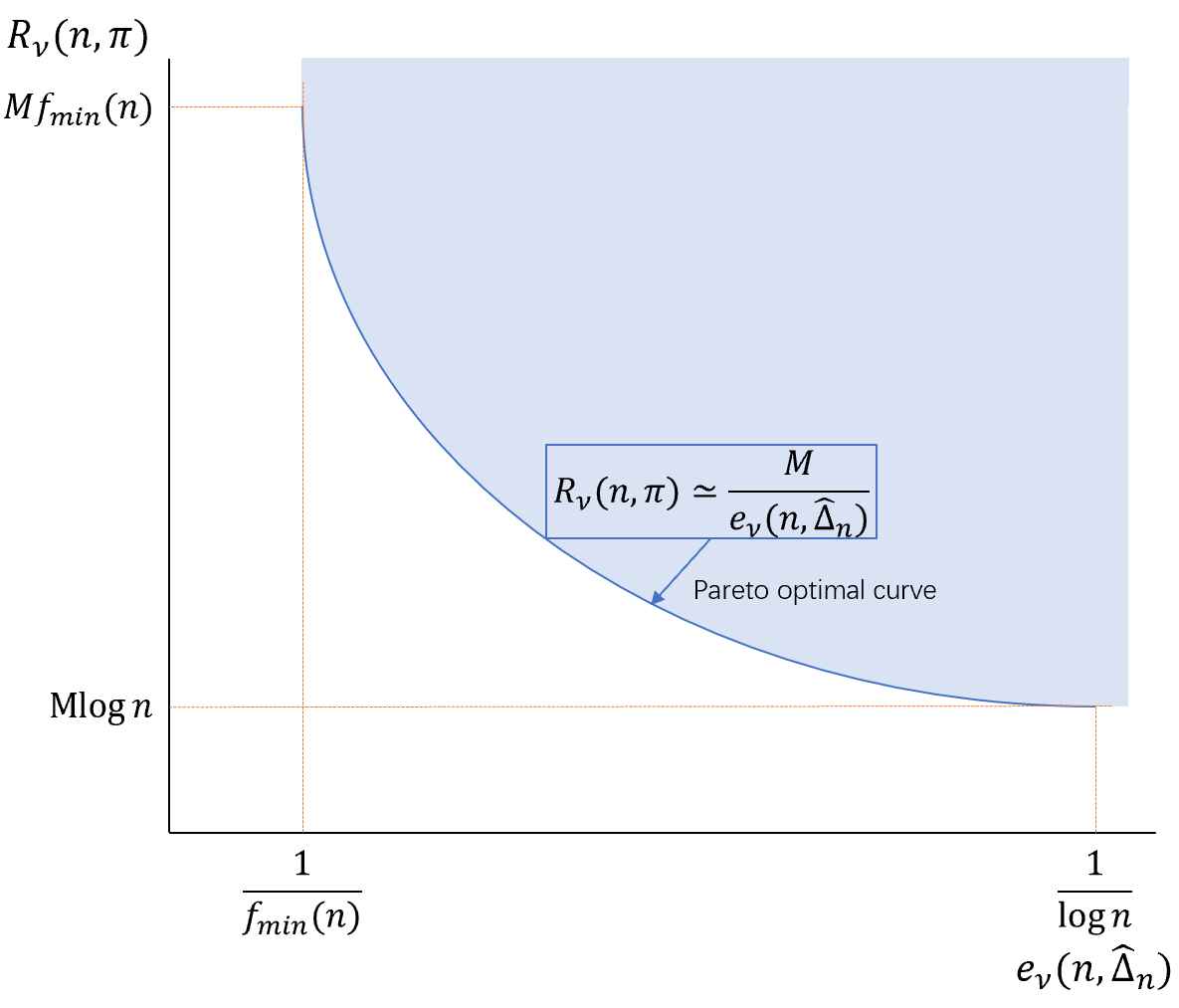}
    \caption{Pareto Optimal Curve}
    \label{fig:noDP}
\end{figure}

\begin{theorem}\label{thm-lower}
    For any admissible pair $\left(\pi, \hat{\Delta}_n\right)$, there always exists a hard instance $\nu \in \mathcal{E}_0$ such that 
    $e_\nu\left(n, \hat{\Delta}_n\right) \ge \Omega\left(\frac{M}{\mathcal{R}_\nu(n, \pi)}\right)$, or in other words
$$
\inf _{\left(\pi, \hat{\Delta}_n\right)} \max _{\nu \in \mathcal{E}_0}\left[e_\nu\left(n, \hat{\Delta}_n\right) \mathcal{R}_\nu(n, \pi)\right] \ge \Omega(M).
$$
\end{theorem}
Theorem \ref{thm-lower} mathematically highlights the trade-off that a small regret will inevitably lead to a large error on the CATE estimation. In specific, it states that for any admissible pair $\left(\pi, \hat{\Delta}_n\right)$, there  exists a hard instance $\nu \in \mathcal{E}$ such that the expected error is lower bounded by $M$ times the inverse of the regret, i.e., $e_\nu\left(n, \hat{\Delta}_n\right) \ge \Omega\left(\frac{M}{\mathcal{R}_\nu(n, \pi)}\right)$. In particular, since $\mathcal{R}_\nu(n, \pi)=\mathcal{O}(\log (n))$ for UCB and TS algorithms, no estimators can not achieve smaller error than the order $\Omega\left(\frac{M}{\log (n)}\right)$ consistently over all the possible instances. Although $\log (n)$ increases with $n$, the speed is rather slow which explicitly shows the limitation of regret optimal policies in terms of statistical power for estimating the CATE. On the other hand, if we ignore the regret and simply run  random control trials (RCT), it can be easily shown that 
$e_\nu(n, \hat{\Delta}_n)= \max_{1\leq j\leq M}\mathbb{E}\left[\left(\hat{\Delta}_n(X_j)-\Delta(X_j)\right)^2\right]=\mathcal{O}\left(\frac{1}{f_{min}(n)}\right)$, which is the best possible accuracy one can obtain but will result in $\mathcal{O}(n)$ regret. The above two cases can be regarded as two extreme cases (note that they don't match the lower bound), but in practice, the experimenter may want to find a balance of estimation accuracy and regret between these two extreme cases. In the following, we provide a family of algorithms called \textbf{ConSE} which depends on a parameter $\alpha \in [0,1]$. A larger $\alpha$ leads to smaller regret and larger estimation error. In particular, when $\alpha=1$, the algorithm ignores estimation error and focuses on minimizing regret. On the contrary, when $\alpha=0$, the algorithm only focuses on minimizing estimation error. Moreover, for each given $\alpha$, \textbf{ConSE} achieves the lower bound provided in theorem \ref{thm-lower}, which shows that it's optimal for every possible cases on the curve from one extreme case to the other (see figure \ref{fig:noDP}). In the figure, the endpoints of the curve represent two extreme case with minimum regret and estimation error. The other points on the curve characterize the tradeoff between these two objectives. Namely, this is the Pareto optimal curve for regret and estimation error. In section \ref{sec:pareto}, we will have a more detailed discussion on variants of the Pareto optimal curve.

In the ConSE algorithm, we need to use the following notation.\\
\textbf{Define three number sequences:}\\
For $e=epoch=1,2,3,...$ and the number of total patients $n$, define:\\
$\Delta_e=2^{-epoch}$\\
$R_e=\max\{\frac{32\log(16n \cdot \text{epoch}^2)}{\Delta_e^2}, \frac{8\log(8n \cdot \text{epoch}^2)}{\Delta_e}\}+1$\\
$h_e=\sqrt{\frac{\log \left(16n \cdot \text{epoch}^2\right)}{2R_e}}$\\
\begin{algorithm}
\caption{ConSE}\label{algorithm1}
\textbf{Input:} $\alpha$.\\
\textbf{Initialize:}$S_j \leftarrow\{0, 1\}$, epoch $e_j$ $\leftarrow 0$, $r_j$ $\leftarrow 0$, $\bar{\mu}_i^j  \leftarrow 0$, $n_j\leftarrow 0$ $(i=0,1; j=1,2,...,M)$.\\
\textbf{for} $t=1,2,...,[\frac{n}{2}]$ \textbf{do}:\\
\quad \textbf{get feature} $x_t=X_{j_t}\in \mathcal{X}$\\
\quad Increment $n_{j_t}\leftarrow n_{j_t}+1$\\
\quad \textbf{if} $|S_{j_t}|=2$:\\
\qquad Select action $a_t\in \{0,1\}$ with equal probabilities $(\frac{1}{2}, \frac{1}{2})$ and update mean $\bar{\mu}_{a_t}^{j_t}$.\\
\qquad Increment $r_{j_t}\leftarrow r_{j_t}+1$\\
\qquad \textbf{if} $r_{j_t} \geq R_{e_{j_t}}$:\\
\qquad \quad \textbf{if} $e_{j_t}\geq 1$:\\
\qquad \qquad Remove arm $i$ from $S_{j_t}$ if
$\max\{\bar{\mu}^{j_t}_1, \bar{\mu}^{j_t}_2\}-\bar{\mu}_i^{j_t}>2 h_e$ $(i=0,1)$\\
\qquad \quad Increment epoch $e_{j_t}\leftarrow e_{j_t}+1$.\\
\qquad \quad Set $r_{j_t}\leftarrow 0$\\
\qquad \quad Zero means: $\bar{\mu}_i^{j_t}\leftarrow 0$ $\forall i\in \{1, 2\}$\\
\quad \textbf{else}:\\
\qquad Pull the arm in $S_{j_t}$.\\
\quad \textbf{if} $t=[\frac{n}{2}]$:\\
\qquad $\hat{f_j}=n_j(1\leq j\leq M)$,\\
\qquad $T_{min}=\max \{\log n, \min \{\hat{f_1}^{1-\alpha}, \hat{f_2}^{1-\alpha}, \cdots, \hat{f_M}^{1-\alpha}\}\}$\\
\textbf{end for}\\
\textbf{for} $j=1,2,...,M$:\\
\quad $n_j=0$\\
\textbf{end for}\\
\textbf{for} $t=[\frac{n}{2}]+1,[\frac{n}{2}]+2,...,n$ \textbf{do}\\
\quad \textbf{get} $x_t=X_{j_t}\in \mathcal{X}$\\
\quad Increment $n_{j_t}\leftarrow n_{j_t}+1$\\
\quad \textbf{if} $n_{j_t}\leq T_{min}$:\\
\qquad Select action $a_t\in \{0,1\}$ with equal probabilities $(\frac{1}{2}, \frac{1}{2})$ and update mean $\bar{\mu}_{a_t}^{j_t}$.\\
\qquad \textbf{if} $n_{j_t}=T_{min}$:\\
\quad \qquad \textbf{Output:} $\hat{\Delta}(X_{j_t})=\bar{\mu}_1^{j_t}-\bar{\mu}_0^{j_t}$\\
\quad \textbf{else:}\\
\qquad Pull the arm in $S_{j_t}$. (if $|S_{j_t}|=2$, pull any arm $a_t\in S_{j_t}$)\\
\textbf{end for}\\
\end{algorithm}
\begin{theorem}\label{theorem-upper}
Let Algorithm 1 runs with any given $\alpha \in[0,1]$. For any instance $\nu$, the regret and estimation error are
$$
\mathcal{R}_\nu(n, \pi)\le \mathcal{O}\left(M\max\{f_{min}(n)^{1-\alpha}, \log n \}\right),
$$
$$
e_\nu(n, \hat{\Delta}_n)= \max_{1\leq j\leq M}\mathbb{E}\left[\left(\hat{\Delta}_n(X_j)-\Delta(X_j)\right)^2\right]\leq \mathcal{O}\left(\frac{1}{\max\{f_{min}(n)^{1-\alpha}, \log n\}}\right).
$$
Therefore, the product of regret and estimation error is always $\mathcal{O}(M)$, i.e.,
$$
e_\nu(n, \hat{\Delta}_n) \mathcal{R}_\nu(n, \pi)\le \mathcal{O}(M)
$$
\end{theorem}

Combining the two theorems above, we can now answer \textit{\textbf{Question 1}}:
Given a budget of social welfare loss $\mathcal{R}_\nu(n, \pi)$, the best possible accuracy of inference for CATE is $\mathcal{O}\left(\frac{M}{\mathcal{R}_\nu(n, \pi)}\right)$ and is attained by \textbf{ConSE}.

\begin{remark}
    Compared to previous work in bandit experiment (\citealt{simchi2023multi}), while the high level idea is similar, we consider an alternative estimator and improve the analysis in the proof. Specifically, in \cite{simchi2023multi}, the upper bound is tight up to poly-log term, while in this paper the upper bound is tight up to constant. First of all, since classical bandit algorithms like UCB or TS attain regret bound of $\mathcal{O}(\log n)$, we believe that poly-log factors do matter. Besides, this improved upper bound help us have a better characterization of Pareto optimal curve that we will explain in section 3. Finally, the estimator in our algorithm is asymptotically normal, which means we can construct (asymptotic)
normal confidence interval for inference and hypothesis testing, which has been a long standing issue for existing adaptive experiment design literature (\citealt{simchi2023multi}, \citealt{zhao2023adaptive}, \citealt{dai2023clip}).
\end{remark}

\begin{table}[t]
\caption{Comparison with \citealt{simchi2023multi}.}
\label{Comparison Table}
\begin{center}
\begin{small}
\begin{sc}
\begin{tabular}{lcccr}
\toprule
Differences &  \citealt{simchi2023multi} &  This Paper \\
\midrule
Context & No& Yes\\
Lower bound & $\Omega(1)$ & $\Omega(M)$ \\
Upper bound & $\mathcal{O}(\log n)$ & $\mathcal{O}(M)$\\
Differential Privacy & No& Yes\\
Asymptotic Normality & No& Yes\\
\bottomrule
\end{tabular}
\end{sc}
\end{small}
\end{center}
\vskip -0.3in
\end{table}

\begin{proposition} \label{CLT 1}
 The estimators for all features $\hat{\Delta}(X_j)$ are unbiased, i.e., $\mathbb{E}\left[\hat{\Delta}(X_j)\right]=\Delta(X_j)$  $(\forall 1\leq j\leq M)$. Moreover, $\hat{\Delta}(X_j)$ is asymptotically normal for any $j$,
 or formally,
 $$ \lim_{n \to \infty} \sqrt{\max\{f_{min}(n)^{1-\alpha}, \log n \}}(\hat{\Delta}(X_j)-\Delta(X_j)) \leadsto \textbf{N}(0, \sigma_{j0}^2+\sigma_{j1}^2 ).$$
\end{proposition}

Intuitively speaking, \textbf{ConSE} can be divided into three steps:

\noindent \textbf{Step 1.} (From line 3 to 20) In the first half periods, we use Successive Elimination algorithm separately for each arm to eliminate the suboptimal arm and maintain $\log n$ regret. At the same time, we use these data to estimate the appearance frequency $f_j(\frac{n}{2})$ of each feature $X_j$ as defined in assumption 1.

\noindent \textbf{Step 2.} (From line 21 to 30) At the beginning of second half periods, we run RCT $\hat{f}_{min} (\frac{n}{2})$ times for every feature, where 
$\hat{f}_{min} (\frac{n}{2})$ is estimated from Phase 1.

\noindent \textbf{Step 3.} (From line 31 to 32)
Play the optimal arm for each feature in the remaining time of experiment.

Although it becomes more complicated with privacy constraints, our main goal is still to do the three steps \textbf{\textit{privately}}. A more detailed discussion will be provided in next section.

\section{Privacy is Free:
A Double-Private Algorithm for Bandit Experiment
}



In this section, our focus is to answer \textbf{\textit{Question 2}}, i.e., \textbf{\textit{with the constraint that the experimenter need to protect the privacy of participants, is it still possible to attain the same estimation accuracy as well as social welfare loss?}} Roughly speaking, our answer is yes (when $\varepsilon$ is a small, constant number, which is the most common case). In other words, we will provide a DP version of \textbf{ConSE} that matches the lower bound provided in theorem \ref{thm-lower} for any given $\alpha \in [0,1]$, where the meaning of $\alpha$ is exactly the same as in \textbf{ConSE} described in last section. The framework of \textbf{DP-ConSE} is quite similar to ConSE, with changes only in technical details.

In the DP-ConSE algorithm, we need to use the following two notations.\\
\textbf{Define a r.v. generator}:\\
Given $\varepsilon>0$, $\forall m>0$, denote $Lap^+(m)=Lap^+_\varepsilon(m)$ is a random variable, satisfies:
$$P(Lap^+(m)=[m]+k)=\frac{e^{-\frac{\varepsilon}{2}|k|}(e^\frac{\varepsilon}{2}-1)}{e^\frac{\varepsilon}{2}+1-e^{\frac{\varepsilon}{2}[m]}} \quad (-[m]\leq k<+\infty, k\in \mathcal{Z})$$
\textbf{Define four number sequences:}\\
For $e=epoch=1,2,3,...$, and $\varepsilon>0$ and the number of total patients $n$, define:\\
$\Delta_e=2^{-epoch}$\\
$R_e=\max\{\frac{32\log(16n \cdot \text{epoch}^2)}{\Delta_e^2}, \frac{8\log(8n \cdot \text{epoch}^2)}{\varepsilon\Delta_e}\}+1$\\
$h_e=\sqrt{\frac{\log \left(16n \cdot \text{epoch}^2\right)}{2R_e}}$\\
$c_e=\frac{2\log \left(8n\cdot \text{epoch}^2\right)}{R_e\varepsilon}$

The proposed algorithm \textbf{DP-ConSE} is formulated as in algorithm \ref{algorithm-dp}. 
As promised, in the following we will give an intuitive description of three steps in DP-ConSE, together with the proof sketches and techniques used to make the algorithm private.

\begin{algorithm}
\caption{DP-ConSE}\label{algorithm-dp}
\textbf{Input:} $\alpha$, privacy-loss $\varepsilon$.\\
\textbf{Initialize:}$S_j \leftarrow\{0, 1\}$, epoch $e_j$ $\leftarrow 0$, $R^j_0=0$, $r_j$ $\leftarrow 0$, $\bar{\mu}_i^j  \leftarrow 0$, $n_j\leftarrow 0$ $(i=0,1; j=1,2,...,M)$.\\
\textbf{for} $t=1,2,...,[\frac{n}{2}]$ \textbf{do}:\\
\quad \textbf{get feature} $x_t=X_{j_t}\in \mathcal{X}$\\
\quad Increment $n_{j_t}\leftarrow n_{j_t}+1$\\
\quad \textbf{if} $|S_{j_t}|=2$:\\
\qquad Select action $a_t\in \{0,1\}$ with equal probabilities $(\frac{1}{2}, \frac{1}{2})$ and update mean $\bar{\mu}_{a_t}^{j_t}$.\\
\qquad Increment $r_{j_t}\leftarrow r_{j_t}+1$\\
\qquad \textbf{if} $r_{j_t} \geq R^{j_t}_{e_{j_t}}$:\\
\qquad \quad \textbf{if} $e_{j_t}\geq 1$:\\
\qquad \qquad Set $\widetilde{\mu}^{j_t}_i \leftarrow \bar{\mu}_i^{j_t}+\operatorname{Lap}(2 / \varepsilon R_{e_{j_t}})$\\
\qquad \qquad Remove arm $i$ from $S_{j_t}$ if
$\max\{\widetilde{\mu}^{j_t}_1, \widetilde{\mu}^{j_t}_2\}-\widetilde{\mu}^{j_t}_i>2 h_e+2 c_e$ $(i=0,1)$\\
\qquad \quad Increment epoch $e_{j_t}\leftarrow e_{j_t}+1$.\\
\qquad \quad Set $r_{j_t}\leftarrow 0$\\
\qquad \quad Set $R^{j_t}_{e_{j_t}}\leftarrow Lap^+_\varepsilon(R_{e_{j_t}})$\\
\qquad \quad Zero means: $\bar{\mu}_i^{j_t}\leftarrow 0$ $\forall i\in \{1, 2\}$\\
\quad \textbf{else}:\\
\qquad Pull the arm in $S_{j_t}$.\\
\quad \textbf{if} $t=[\frac{n}{2}]$:\\
\qquad $\hat{f_j}=n_j(1\leq j\leq M)$,\\
\qquad $T_{min}=\max \{\frac{\log n}{\varepsilon}, \min \{\hat{f_1}^{1-\alpha}, \hat{f_2}^{1-\alpha}, \cdots, \hat{f_M}^{1-\alpha}\}\}$\\
\textbf{end for}\\
\textbf{for} $j=1,2,...,M$:\\
\quad $T_j=Lap_\varepsilon^+(T_{min})$\\
\quad $n_j=0$\\
\textbf{end for}\\
\textbf{for} $t=[\frac{n}{2}]+1,[\frac{n}{2}]+2,...,n$ \textbf{do}\\
\quad \textbf{get} $x_t=X_{j_t}\in \mathcal{X}$\\
\quad Increment $n_{j_t}\leftarrow n_{j_t}+1$\\
\quad \textbf{if} $n_{j_t}\leq T_{j_t}$:\\
\qquad Select action $a_t\in \{0,1\}$ with equal probabilities $(\frac{1}{2}, \frac{1}{2})$ and update mean $\bar{\mu}_{a_t}^{j_t}$.\\
\qquad \textbf{if} $n_{j_t}=T_{j_t}$:\\
\quad \qquad \textbf{Output:} $\hat{\Delta}(X_{j_t})=\bar{\mu}_1^{j_t}-\bar{\mu}_0^{j_t}+Lap(2/\varepsilon T_{j_t})$\\
\quad \textbf{else:}\\
\qquad Pull the arm in $S_{j_t}$. (if $|S_{j_t}|=2$, pull any arm $a_t\in S_{j_t}$)\\
\textbf{end for}\\
\end{algorithm}

\noindent \textbf{Step 1.}  In the first half periods, we use an improved "DP Successive Elimination" algorithm in \cite{sajed2019optimal} for each feature. Our goal in this phase is twofold for each feature: to identify the optimal action and estimate the frequency of occurrences (based on our non-stationary seasonal assumption) with minimal regret. For each feature we compare the \textbf{privatized} average rewards of two actions in batches. If the difference is large, we eliminate the sub-optimal arm and claim that we find the optimal arm with high probability. There are two technical designs involved here. First, the length of batches increases exponentially, which strikes a balance between differential privacy protection and regret loss. Similar idea can be found in “DP Successive Elimination” algorithm (\citealt{sajed2019optimal}) and widely used "tree mechanism" (\cite{chan2011private}) in DP-bandit algorithms. Second, we use a novel technique by adding noise to the batch lengths for each feature. The reason for this is that due to the seasonal non-stationarity  assumption, it's essential to run batched learning for each feature independently, and to protect the patients' feature, the length of batches should also be privatized.
To the best of our knowledge, this technique has not appeared in DP-bandit literature and again highlights the difficulty of DP-contextual bandit compared to bandit setting. 

After identifying the optimal action, we will continue to execute this action until the first half of the experiment is completed. After the completion of the first half, based on the occurrence frequencies of features observed, we can
estimate $f_j(n)$ for each feature $X_j$. This helps us to decide the  length of RCTs in second half periods to estimate
CATE.

To make our claim valid, we first need to show that the elimination process will end in \textbf{step 1} (with high probability). This is confirmed by the following lemma. 
\begin{lemma} \label{lemma 1}
Let DP-ConSE runs with any given $\alpha \in[0,1]$ and $\varepsilon>0$. Then w.p. $\geq 1-\frac{1}{n}$ it holds that DP-ConSE pulls the bad arm of any feature $X_j$ in the first half periods for at most
$$
\mathcal{O}\left(\left(\log n_j+\log \log \left(1 / \Delta(X_j)\right)\right)\left(\frac{1}{\Delta(X_j)^2}+\frac{1}{\varepsilon \Delta(X_j)}\right)\right)
$$
where $n_j$ is the number of occurrences of the feature $X_j$ $(1\leq j\leq M)$.
\end{lemma}

So when $\varepsilon$ is a small constant and $n$ is sufficiently large, we can find the optimal arm for each feature in the first half with high probability, and the number of playing suboptimal arm is bounded. As a corollary, we can bound the regret in the first half periods as claimed.
\begin{corollary} \label{corollary 1}
    For sufficiently large $n$, the expected pseudo regret in the first half periods of DP-ConSE is at most $\mathcal{O}\left(\left(\sum_{1 \leq j\leq M} \frac{\log n}{\Delta(X_j)}\right)+\frac{M\log n}{\varepsilon}\right)$.
\end{corollary}

\noindent \textbf{Step 2.} In the second half periods, our primary objective is to ensure the required accuracy of estimating the CATE. Using the estimated $f_j(n)$ from \textbf{step 1}, we can determine the length of RCTs for each feature to attain the desired accuracy. It is important to remember that we still need to add noise to the length of RCTs for the same reason as stated in \textbf{step 1}.

After \textbf{step 2}, the main task of estimating CATE is completed, and the estimation accuracy is provided in the following theorem.
\begin{theorem} \label{theorem 3}
    If DP-ConSE runs with $\alpha \in[0,1]$ and $\varepsilon>0$, the estimate error is
$$
e(n, \hat{\Delta})=\mathcal{O}\left(\frac{1}{\max\{f_{min}(n)^{1-\alpha}, \frac{\log n}{\varepsilon}\}}\right).
$$
\end{theorem}

\noindent \textbf{Step 3.}  Finally, for each feature, after completing RCT phase in \textbf{step 2}, we simply play the optimal action obtained in the first half periods for the remaining patients with the aim of achieving minimum regret. The cumulative regret in the second half periods can be bounded as in the following lemma.
\begin{lemma} \label{lemma 2}
    The expected regret in the second half periods of DP-ConSE is at most 
    $$\mathcal{O}\left(\max\{f_{min}(n)^{1-\alpha}, \frac{\log n}{\varepsilon} \}\sum_{1 \leq j\leq M} \Delta(X_j)\right).$$
\end{lemma}
We have elaborated on how our algorithm strikes a balance between estimation, regret minimization and differential privacy, and to wrap things up, we have the following theorem to answer \textbf{\textit{Question 2}}. A rigorous proof can be found in appendix.
\begin{theorem}\label{theorem-dpupper}
   DP-ConSE is $(\varepsilon, \frac{1}{n})$-private. Moreover, let DP-ConSE runs with any given $\alpha \in[0,1]$ and $\varepsilon>0$. The regret is
$$
\mathcal{O}\left(M\max\{f_{min}(n)^{1-\alpha}, \frac{\log n}{\varepsilon} \}\right),
$$
As a result, we have
$$
e_\nu(n, \hat{\Delta}_n) \mathcal{R}_\nu(n, \pi)\le \mathcal{O}(M),
$$
which is the same as theorem \ref{theorem-upper}
and matches the lower bound in theorem \ref{thm-lower}.
\end{theorem}

Similar to the estimator without privacy constraint, we can also prove that the estimator in DP-ConSE is asymptotically normal, and more interestingly, it has the same asymptotic variance as in ConSE. This
again shows that privacy is "free" in the case of statistical inference, as the Laplacian noise converges to 0 faster as $n \to \infty$ compared to Guassian variable, and thus has no impact on asymptotic variance of the estimator.
\begin{proposition} \label{CLT 2}
    The estimators for all features $\hat{\Delta}(X_j)$ are unbiased, i.e., $\mathbb{E}\left[\hat{\Delta}(X_j)\right]=\Delta(X_j)$  $(\forall 1\leq j\leq M)$. Moreover, $\hat{\Delta}(X_j)$ is asymptotically normal for any $j$,
 or formally,
 $$ \lim_{n \to \infty} \sqrt{\max\{f_{min}(n)^{1-\alpha}, \frac{\log n}{\varepsilon} \}}(\hat{\Delta}(X_j)-\Delta(X_j)) \leadsto \textbf{N}(0, \sigma_{j0}^2+\sigma_{j1}^2 ).$$
\end{proposition} 

\section{Pareto Optimal Curve} \label{sec:pareto}

In this section, we will characterize the Pareto optimal curve of regret and estimation error, which is the standard measurement in multi-objective optimization problems and is widely adopted in multi-objective bandit literature (\citealt{simchi2023multi}, \citealt{zhong2021achieving}). A formal definition is provided in the following. 
\begin{definition}
    A pair of regret and estimation error $(x(n),y(n))$ is \textbf{\textit{Pareto optimal}} with respect to $n$ if there exists no algorithm which can attain a regret and estimation error pair $(\alpha(n), \beta(n))$ such that 
    $\alpha(n) \le \mathcal{O}(x(n)) $, $\beta (n)=o(y(n)) $ or 
     $\alpha(n) = o(x(n)) $, $\beta (n)\le\mathcal{O}(y(n)) $ as $n \to \infty$. 
\end{definition}

From the upper and lower bound we derive above in theorem \ref{thm-lower}, \ref{theorem-upper}, \ref{theorem-dpupper}, we know exactly what the Pareto optimal curve is.
\begin{theorem}
    The Pareto optimal curve for regret and estimation error (with or without privacy constraint) is characterized by
    $$
e_\nu(n, \hat{\Delta}_n) \mathcal{R}_\nu(n, \pi)= \mathcal{O}(M),
$$
\end{theorem}

\begin{figure}
    \centering
    \includegraphics[width=0.5\linewidth]{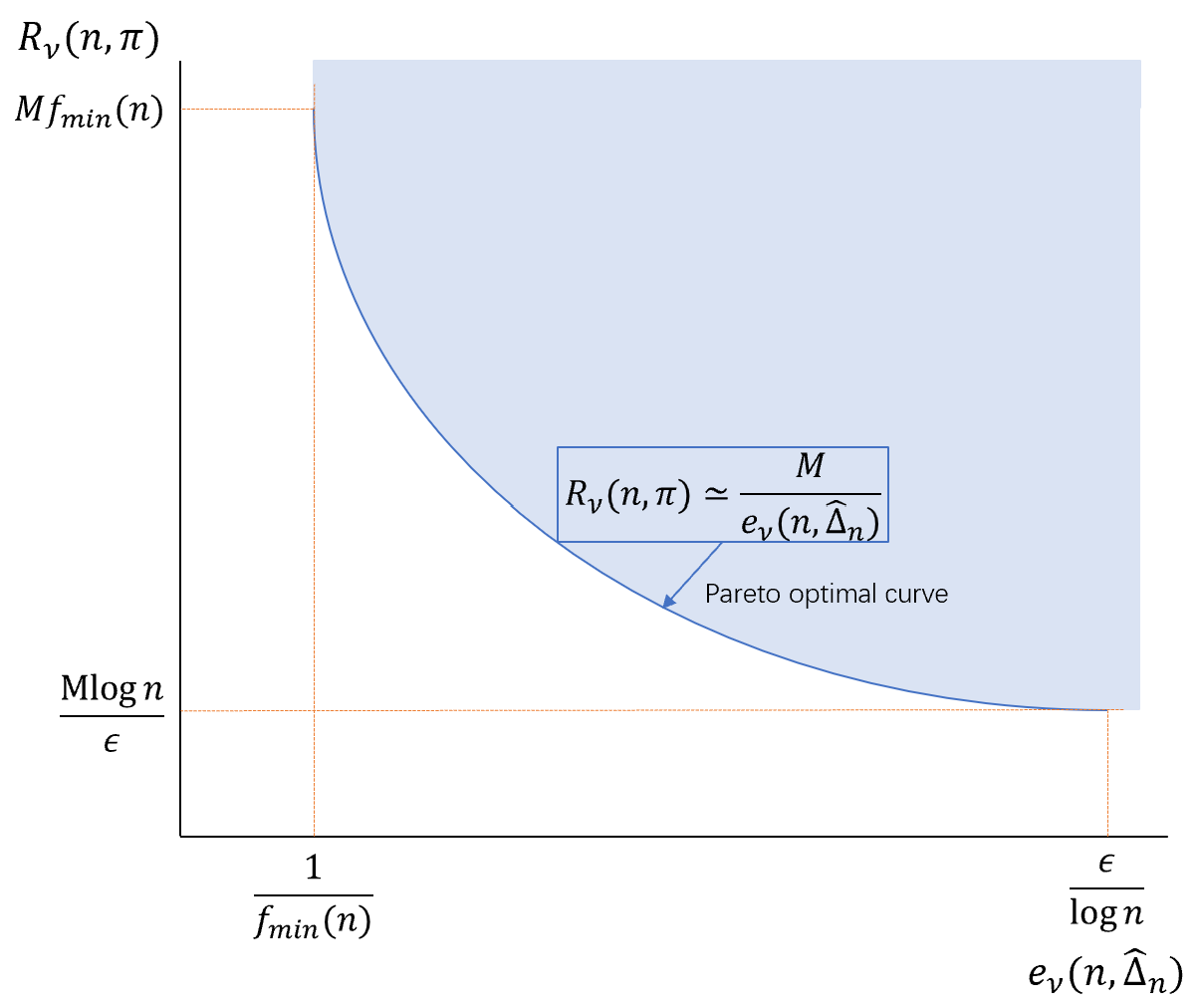}
    \caption{Pareto Optimal Curve (General)}
    \label{fig:generalDP}
\end{figure}

Figure \ref{fig:generalDP} shows the Pareto optimal curve in general case. We can see that the applicability of the Pareto optimal curve depends on the privacy protection indicator $\epsilon$ and the minimum feature occurrence $f_{min}$. As for the privacy protection parameter $\epsilon$, we claim that it is almost "for free" in terms of the trade-off between regret and estimation error, as it does not affect the equation of our Pareto optimal curve. Here we can clearly see that the only cost of privacy protection is that the higher the requirement for privacy protection, the shorter our optimal curve will be. This is due to the fact that the smallest possible estimation error and regret depend on $\varepsilon$, which is $ \frac{\epsilon}{\log n}$and $\frac{M \log n}{\varepsilon}$, respectively.

\begin{figure}
    \centering
    \includegraphics[width=0.5\linewidth]{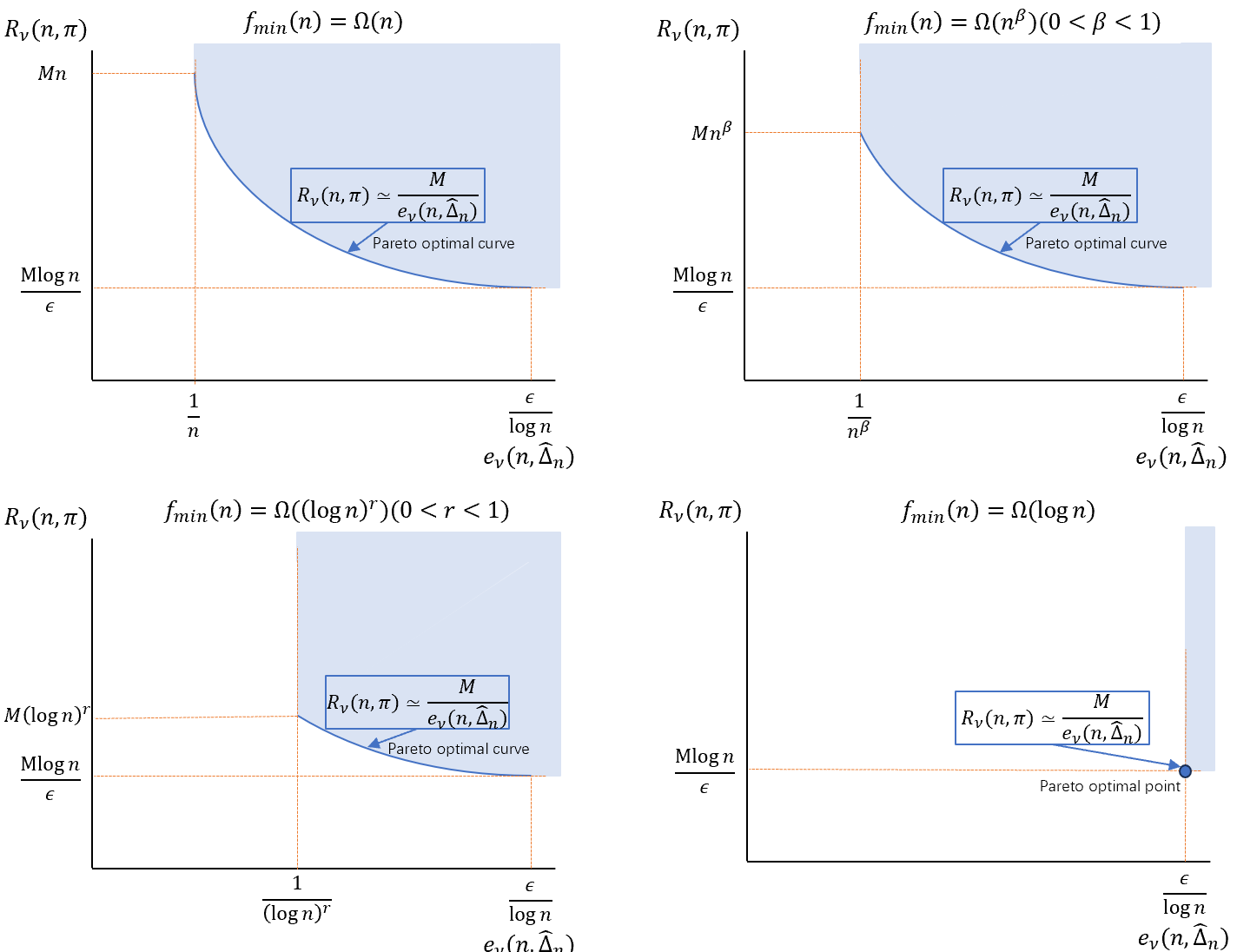}
    \caption{Pareto Optimal Curves (In Different Scenarios)}
    \label{fig:example}
\end{figure}

The minimum feature occurrence $f_{min}$ is entirely determined by the (nonstationary) distribution of patient features rather than by the experimenter. Therefore, in different scenarios, we may have different optimal curves.
Figure \ref{fig:example} shows Pareto optimal curves in different scenarios, namely when $f_{min}$ takes different values. We can see that as $f_{min}$ decreases gradually, the Pareto optimal curve shortens and eventually collapses into a single point. Compared to \cite{simchi2023multi}, since they cannot deal with sub-polynomial terms, it's impossible for them to characterize the case when $f_{min}= \text{polylog} (n)$ (see figure \ref{fig:example}).   The light blue areas in different scenarios intuitively show the regions of the pair $\left(e_\nu(n, \hat{\Delta}_n),\mathcal{R}_\nu(n, \pi)\right)$ that different algorithms may achieve, while there exists no algorithm that can attain the region below Pareto optimal curves. 

\section{Concluding Remarks}

In this paper, we statistically investigate the trade-off between efficiency in decision-making and estimation precision
of CATE in contextual bandit experiments. We adopt the minimax multi-objective optimization framework
and Pareto optimality to characterize the trade-off. We first provide a lower bound of the multi-objective optimization problem and then propose ConSE to match that lower bound. Going one step further, we consider the constraint of protecting patients' privacy and propose a differentially private version of ConSE (DP-ConSE) which still matches the lower bound, demonstrating that privacy is "almost" free. Besides, we also develop the asymptotic normality for both ConSE and DP-ConSE, which is crucial for statistical inference and hypothesis testing.

\printendnotes

\bibliographystyle{chicago}
\bibliography{citation}

\section{Appendix}
\subsection{Proof to Theorem \ref{thm-lower}}
We first consider the case that $M=1$. In this case, we are actually talking about an ATE problem, that means no feature. And first, we start from proving the Lemma 1.1 below.\\
   \textbf{Lemma 1.1} For any given online decision-making policy $\pi$, the error of any ATE estimators can be lower bounded as follows, for any function $\phi: n \rightarrow\left[0, \frac{1}{4}\right]$ and any $u \in \mathcal{E}_0$
$$
\inf _{\hat{\Delta}_n} \max _{\nu \in \mathcal{E}_0} \mathbb{P}_\nu\left(\left|\hat{\Delta}_n-\Delta_\nu\right| \geq \phi(n)\right) \geq \frac{1}{2}\left[1-\sqrt{\frac{16}{3} \phi(n)^2 \frac{\mathcal{R}_u(n, \pi)}{\left|\Delta_u\right|}}\right]
$$
We define a Rademacher-like distribution as $X \sim \operatorname{Rad}(p)$ means $X=-1$ with probability $p$ and $X=1$ with probability $1-p$. Consider the following two bandits instance $\nu_1=$ $\left(\operatorname{Rad}\left(\frac{1-\xi}{2}\right), \operatorname{Rad}\left(\frac{1}{2}\right)\right)$ and $\nu_2=\left(\operatorname{Rad}\left(\frac{1-\xi}{2}\right), \operatorname{Rad}\left(\frac{1+2 \phi(t)}{2}\right)\right)$. Note that the treatment effects of $\nu_1$ and $\nu_2$ is $\Delta_1=\xi$ and $\Delta_2=\xi+2 \phi(t)$. $\xi$ can be any number in $(0,1]$. By such constructions and the symmetry, $\nu_1$ and $\nu_2$ can represent all the possible instances without loss of generality. We define the minimum distance test $\psi\left(\hat{\Delta}_t\right)$ that is associated to $\hat{\Delta}_t$ by $\psi\left(\hat{\Delta}_t\right)=\arg \min _{i=1,2}\left|\hat{\Delta}_t-\Delta_i\right|$. If $\psi\left(\hat{\Delta}_t\right)=1$, we know that $\left|\hat{\Delta}_t-\Delta_1\right| \leq\left|\hat{\Delta}_t-\Delta_2\right|$. By the triangle inequality, we can have, if $\psi\left(\hat{\Delta}_t\right)=1$,
$$
\left|\hat{\Delta}_t-\Delta_2\right| \geq\left|\Delta_1-\Delta_2\right|-\left|\hat{\Delta}_t-\Delta_1\right| \geq\left|\Delta_1-\Delta_2\right|-\left|\hat{\Delta}_t-\Delta_2\right|,
$$
which yields that $\left|\hat{\Delta}_t-\Delta_2\right| \geq \frac{1}{2}\left|\Delta_1-\Delta_2\right|=\phi(t)$. Symmetrically, if $\psi\left(\hat{\Delta}_t\right)=2$, we can have $\mid \hat{\Delta}_t-$ $\Delta_1\left|\geq \frac{1}{2}\right| \Delta_1-\Delta_2 \mid=\phi(t)$. Therefore, we can use this to show

\begin{eqnarray}
\inf _{\hat{\Delta}_t} \max _{\nu \in \mathcal{E}_0} \mathbb{P}_\nu\left(\left|\hat{\Delta}_t-\Delta_\nu\right|_2 \geq \phi(t)\right) &\geq& \inf _{\hat{\Delta}_t} \max _{i \in\{1,2\}} \mathbb{P}_{\nu_i}\left(\left|\hat{\Delta}_t-\Delta_i\right|_2 \geq \phi(t)\right)\nonumber \\
&\geq& \inf _{\hat{\Delta}_t} \max _{i \in\{1,2\}} \mathbb{P}_{\nu_i}\left(\psi\left(\hat{\Delta}_t\right) \neq i\right)\nonumber \\
&\geq& \inf _\psi \max _{i \in\{1,2\}} \mathbb{P}_{\nu_i}(\psi \neq i)\nonumber
\end{eqnarray}

where the last infimum is taken over all tests $\psi$ based on $\mathcal{H}_t$ that take values in $\{1,2\}$.

\begin{eqnarray}
\inf _{\hat{\Delta}_t} \max _{\nu \in \mathcal{E}_0} \mathbb{P}_\nu\left(\left|\hat{\Delta}_t-\Delta_\nu\right|_2 \geq \phi(t)\right) &\geq& \inf _\psi \max _{i \in\{1,2\}} \mathbb{P}_{\nu_i}(\psi \neq i)\nonumber \\
&\geq& \frac{1}{2} \inf _\psi\left(\mathbb{P}_{\nu_1}(\psi=2)+\mathbb{P}_{\nu_2}(\psi=1)\right)\nonumber \\
&=&\frac{1}{2}\left[1-\operatorname{TV}\left(\mathbb{P}_{\nu_1}, \mathbb{P}_{\nu_2}\right)\right]\nonumber \\
&\geq& \frac{1}{2}\left[1-\sqrt{\frac{1}{2} \mathrm{KL}\left(\mathbb{P}_{\nu_1}, \mathbb{P}_{\nu_2}\right)}\right]\nonumber \\
&\geq& \frac{1}{2}\left[1-\sqrt{\frac{8 \phi(t)^2}{3 \xi} \mathcal{R}_{\nu_1}(t, \pi)}\right]\nonumber
\end{eqnarray}

where the equality holds due to Neyman-Pearson lemma and the third inequality holds due to Pinsker's inequality, and the fourth inequality holds due to the following:

\begin{eqnarray}
\mathrm{KL}\left(\mathbb{P}_{\nu_1}, \mathbb{P}_{\nu_2}\right) &=&\sum_{s=1}^t \mathbb{E}_{\nu_1}\left[\operatorname{KL}\left(P_{1, A_t}, P_{2, A_t}\right)\right]\nonumber\\
&=&\sum_{i=1}^2 \mathbb{E}_{\nu_1}\left[T_i(n)\right] \operatorname{KL}\left(P_{1, i}, P_{2, i}\right)\nonumber \\
&=&\frac{(\phi(t))^2}{\frac{1}{4}-(\phi(t))^2}\left(\mathbb{E}_{\nu_1}\left[T_i(n)\right]\right)\nonumber\\
&\leq& \frac{16 \phi(t)^2}{3 \xi} \mathcal{R}_{\nu_1}(t, \pi)\nonumber
\end{eqnarray}
where we use $\operatorname{KL}\left(\operatorname{Rad}\left(\frac{1}{2}\right), \operatorname{Rad}\left(\frac{1+2 \phi(t)}{2}\right)\right)=\frac{\phi(t)^2}{1 / 4-\phi(t)^2} \leq \frac{16 \phi(t)^2}{3}$, and the last inequality holds because the history $\mathcal{H}_t$ is generated by $\pi$ and $\xi \mathbb{E}_{\nu_1}\left[T_i(n)\right]$ is just the expected regret of $\nu_1$, which is just the definition of regret. Now we finish the proof of Lemma 1.1.\\
Then, we can have, given policy $\pi$, and $\hat{\Delta}_n$, if $\phi(n) \leq \sqrt{\frac{3\left|\Delta_u\right|}{32 \mathcal{R}_u(n, \pi)}}$ for some $u \in \mathcal{E}_0$,
\begin{eqnarray}
\max _{\nu \in \mathcal{E}_0} \mathbb{E}\left[\left|\hat{\Delta}_n-\Delta_\nu\right|^2\right]&\geq&
\phi(t)^2 \max _{\nu \in \mathcal{E}_0} \mathbb{P}_\nu\left(\left|\hat{\Delta}_t-\Delta_\nu\right|_2 \geq \phi(t)\right)\nonumber \\ 
&\geq& \frac{\phi(n)^2}{2}\left[1-\sqrt{\frac{8 \phi(t)^2}{3 \xi} \mathcal{R}_{\nu_1}(t, \pi)}\right]\nonumber \\
&\geq& \frac{\phi(n)^2}{4},
\end{eqnarray}
where the second inequality holds due to Lemma 1.1.\\
We use $\nu_{\pi, \hat{\Delta}_n}$ to denote $\arg \max _{\nu \in \mathcal{E}_0} \mathbb{E}\left[\left|\hat{\Delta}_n- \Delta_\nu\right|^2\right]$ for any given policy $\pi$ and $\hat{\Delta}_n$,
\begin{eqnarray}
\max _{\nu \in \mathcal{E}_0}\left[e_\nu\left(n, \hat{\Delta}_n\right) \mathcal{R}_\nu(n, \pi)\right]
&\geq& e_{\nu_{\pi, \hat{\Delta}_n}}\left(n, \hat{\Delta}_n\right) \mathcal{R}_{\nu_{\pi, \hat{\Delta}_n}}(n, \pi) \nonumber \\
&\geq& \frac{\phi(n)^2}{4} \mathcal{R}_{\nu_{\pi, \hat{\Delta}_n}}(n, \pi)\nonumber \\
&=&\Theta(1),
\end{eqnarray}
where the last equation holds because we plug in $\phi(n)$ and $\Delta_\nu=\Theta(1)$ for $\nu \in \mathcal{E}_0$. Since the above inequalities hold for any policy $\pi$ and $\hat{\Delta}_n$, we finish the proof of the no feature case.\\
In general case, for any $1\leq j \leq M$, we have the following:\\
\begin{eqnarray}
\mathcal{R}_\nu^j(n, \pi)&:=&\mathbb{E}^\pi\left[\sum_{i=1}^n I_{\{x_i=X_j\}}\left[r_i(a^*(x_i)|x_i)- r_i\left(a_i|x_i\right)\right]\right]\\
&=&\mathbb{E}\left[\mathcal{R}_\nu(n_j, \pi)\right], 
\end{eqnarray}
where $n_j=\sum_{i=1}^nI_{\{x_i=X_j\}}$ is a random variable and $\mathbb{E}\left[\sum_{j=1}^M\mathcal{R}_\nu(n_j, \pi)\right]=\sum_{j=1}^M\mathcal{R}_\nu^j(n, \pi)=\mathcal{R}_\nu(n, \pi)$.\\
For using the result in no feature case, consider the following two bandits instance $\nu_1^M=\{X_j:\nu_1|1\leq j\leq M\}$, $\nu_2^M=\{X_j:\nu_2|1\leq j\leq M\}$, where $\nu_1=$ $\left(\operatorname{Rad}\left(\frac{1-\xi}{2}\right), \operatorname{Rad}\left(\frac{1}{2}\right)\right)$ and $\nu_2=\left(\operatorname{Rad}\left(\frac{1-\xi}{2}\right), \operatorname{Rad}\left(\frac{1+2 \phi(t)}{2}\right)\right)$ are introduced in the case $M=1$.\\
Using the result in no feature case, we have the following:\\
\begin{eqnarray}
\max _{\nu \in \mathcal{E}_0}\left[e_\nu\left(n, \hat{\Delta}_n(X_j)\right) \mathcal{R}_\nu(n_j, \pi)\right] \geq \Theta(1)
\end{eqnarray}
for any given $n_j$ and $1\leq j\leq M$.\\
Add their squares up, for any given $\{n_j\}_{j=1}^M$,we have\\
\begin{eqnarray}
\max _{\nu \in \mathcal{E}_0}\left[\left(\max_{1\leq j\leq M}e_\nu\left(n, \hat{\Delta}_n(X_j)\right) \right)\sum_{j=1}^M\mathcal{R}_\nu(n_j, \pi)\right] 
\geq \Theta(M),
\end{eqnarray}
Therefore, we have the result\\
\begin{eqnarray}
\max _{\nu \in \mathcal{E}_0}\left[\left(\max_{1\leq j\leq M}e_\nu\left(n, \hat{\Delta}_n(X_j)\right) \right)\mathcal{R}_\nu(n, \pi)\right] 
\geq \Theta(M),
\end{eqnarray}
Q.E.D.
\subsection{Proof to Theorem \ref{theorem-upper}}
Firstly, we give the proof of $\mathcal{R}_\nu(n, \pi)\le \mathcal{O}\left(M\max\{f_{min}(n)^{1-\alpha}, \log n \}\right)$ below.\\
\textbf{Lemma 2.1} Let Algorithm 1 runs with any given $\alpha \in[0,1]$. Then w.p. $\geq 1-\frac{1}{n}$ it holds that Algorithm 1 pulls the bad arm of any feature $X_j$ in the first half periods for at most
$$
\mathcal{O}\left(\left(\log n_j+\log \log \left(1 / \Delta(X_j)\right)\right)\frac{1}{\Delta(X_j)^2}\right)
$$
where $n_j$ is the number of occurrences of the feature $X_j$ $(1\leq j\leq M)$.\\
\textbf{Proof of Lemma 2.1}\\
Given an epoch $e$ we denote by $\mathcal{E}_e$ the event where for all arms $a \in S$ it holds that $\left|\mu_a-\bar{\mu}_a\right| \leq h_e$ and also denote $\mathcal{E}=\bigcap_{e \geq 1} \mathcal{E}_e$. (we use $T:=n_j$ represents the number of occurrences of the feature $X_j$ and $\beta=\frac{1}{n}$ below)\\
First, by definition, we can calculate that:\\
$R_1\geq 16\log T$, so $R_e\geq 2R_{e-1}\geq 2^{e+3}\log T$.\\
Furthermore, the Hoeffding bound gives that $\operatorname{Pr}\left[\mathcal{E}_e\right] \geq 1-\frac{\beta}{4 e^2}$, thus $\operatorname{Pr}[\mathcal{E}] \geq 1-\frac{\beta}{4}\left(\sum_{e \geq 1} e^{-2}\right)\geq 1-\frac{1}{T}$ ($T\geq 3$). The remainder of the proof continues under the assumption the $\mathcal{E}$ holds, and so, for any epoch $e$ and any viable arm $a$ in this epoch we have $\left|\mu_a-\bar{\mu}_a\right| \leq h_e$. As a result for any epoch $e$ and any two arms $a^1, a^2$ we have that $\left|\left(\bar{\mu}_{a^1}-\bar{\mu}_{a^2}\right)-\left(\mu_{a^1}-\mu_{a^2}\right)\right| \leq 2 h_e$.

Next, we argue that under $\mathcal{E}$ the optimal arm $a^*$ is never eliminated. Indeed, for any epoch $e$, we denote the arm $a_e=\operatorname{argmax}_{a \in S} \bar{\mu}_a$ and it is simple enough to see that $\bar{\mu}_{a_e}-\bar{\mu}_{a^*} \leq$ $0+2 h_e$, so the algorithm doesn't eliminate $a^*$.

Next, we argue that, under $\mathcal{E}$, in any epoch $e$ we eliminate all viable arms with suboptimality gap $\geq 2^{-e}=\Delta_e$. Fix an epoch $e$ and a viable arm $a$ with suboptimality gap $\Delta_a \geq \Delta_e$. Note that we have set parameter $R_e$ so that
$$
\begin{aligned}
h_e & =\sqrt{\frac{\log \left(16 \cdot e^2 / \beta\right)}{2 R_e}}<\sqrt{\frac{\log \left(16 \cdot e^2 / \beta\right)}{2 \cdot \frac{32 \log \left(16 e^2 / \beta\right)}{\Delta_e^2}}}=\frac{\Delta_e}{8};
\end{aligned}
$$

Therefore, since arm $a^*$ remains viable, we have that $\bar{\mu}_{\max }-\bar{\mu}_a \geq \bar{\mu}_{a^*}-\bar{\mu}_a \geq \Delta_a-\left(2 h_e\right)>$ $\Delta_e\left(1-\frac{2}{8}-\frac{2}{8}\right) \geq \frac{\Delta_e}{2}>2 h_e$, guaranteeing that arm $a$ is removed from $S$.

Lastly, fix a suboptimal arm $a$ and let $e(a)$ be the first epoch such that $\Delta_a \geq \Delta_{e(a)}$, implying $\Delta_{e(a)} \leq \Delta_a<\Delta_{e(a)-1}=2 \Delta_e$. Using the immediate observation that for any epoch $e$ we have $R_e \leq R_{e+1} / 2$, we have that the total number of pulls of arm $a$ is
$$
\sum_{e \leq e(a)} R_e \leq \sum_{e \leq e(a)} 2^{e-e(a)} R_{e(a)} \leq R_{e(a)} \sum_{i \geq 0} 2^{-i} \leq 6\left(\frac{32 \log \left(16 \cdot e(a)^2 / \beta\right)}{\Delta_e^2}+\frac{8 \log \left(8 \cdot e(a)^2 / \beta\right)}{\Delta_e}\right)
$$

The bounds $\Delta_e>\Delta_a / 2,|S| \leq 2, e(a)<\log _2\left(2 / \Delta_a\right)$ allow us to conclude and infer that under $\mathcal{E}$ the total number of pulls of arm $a$ is at most 
\begin{equation}
    \log \left(2 \log \left(2 / \Delta_a\right) / \beta\right)\left(\frac{1024}{\Delta_a^2}+\frac{96}{\Delta_a}\right)=\mathcal{O}\left(\left(\log n_j+\log \log \left(1 / \Delta(X_j)\right)\right)\frac{1}{\Delta(X_j)^2}\right) \nonumber
\end{equation}

We finish the proof of Lemma 2.1.\\
Therefore we have the following straightforward corollary.\\
\textbf{Corollary 2.2} For sufficiently large $n$, the expected pseudo regret in the first half periods of Algorithm 1 is at most $\mathcal{O}\left(\sum_{1 \leq j\leq M} \frac{\log n}{\Delta(X_j)}\right)$.\\
Actually, by using the result of lemma 2.1, for $n>\log 1/\Delta(X_j)$, we have
\begin{equation*}
\begin{aligned}
    \mathcal{R}^{first}_\nu(n, \pi)&\leq \Sigma_{1\leq j\leq M}\Delta(X_j)\mathcal{O}\left(\left(\log n_j+\log \log \left(1 / \Delta(X_j)\right)\right)\frac{1}{\Delta(X_j)^2}\right)\\
    &\leq \mathcal{O}\left(\sum_{1 \leq j\leq M} \frac{\log n}{\Delta(X_j)}\right)
\end{aligned}
\end{equation*}\\
For the regret of the second half periods, noticed that with the probability $\geq 1-\frac{1}{n}$, the optimal arm would be chosen correctly in the first half periods. Therefore, the expected regret of the second half periods of algorithm 2 is:\\
\begin{equation}
\mathcal{R}^{second}_\nu(n, \pi)\leq \Sigma_{1\leq j\leq M}\Delta(X_j)\mathbb{E}\left[T_{min}\right]+\frac{1}{n}\mathcal{O}\left(n\right)=\mathcal{O}\left(\max\{f_{min}(n)^{1-\alpha}, \log n \}\sum_{1 \leq j\leq M} \Delta(X_j)\right) \nonumber
\end{equation}
Therefore, when $\Delta(X_j)=\mathcal{O}(1)$ $(\forall 1\leq j\leq M)$, we have
\begin{equation}
    \mathcal{R}_\nu(n, \pi)=\mathcal{R}^{first}_\nu(n, \pi)+\mathcal{R}^{second}_\nu(n, \pi)\le \mathcal{O}\left(M\max\{f_{min}(n)^{1-\alpha}, \log n \}\right) \nonumber
\end{equation}
Secondly, we give the proof of $e_\nu(n, \hat{\Delta}_n)\leq \mathcal{O}\left(\frac{1}{\max\{f_{min}(n)^{1-\alpha}, \log n\}}\right)$ below.\\
Note that for any feature $X_j(1\leq j\leq M)$, we learn at least $T_j=\min\{f_j(n)-f_j(\frac{n}{2}), T_{min}\}$ periods with equal probabilities of two arms, so the MSE estimation error of feature $X_j$ is bounded as $\mathcal{O}\left(\frac{1}{T_j}\right)$. Hence, our estimation error is bounded as $\mathcal{O}\left(\frac{1}{\min_{1\leq j\leq M}{T_j}}\right)$.\\
Now we focus on $T_j$.\\
For any $1\leq j\leq M$ and $1\leq t\leq n$, notice the characteristic function $I_{\{x_t=X_j\}}\in \{0,1\}$ and follows Bernoulli($p_j^t$), we have $\mathbb{E}\left[\hat{f_j}\right]=f_j(\frac{n}{2})=\sum_{1\leq t\leq \frac{n}{2}}p_j^t$\\
Therefore, by Chernoff bound (multiplicative form (relative error)), we have\\
\begin{eqnarray}
\mathbb{P}\left(\hat{f_j}<\frac{f_j(\frac{n}{2})}{2}\right)
&\leq& \left(\frac{e^{-\frac{1}{2}}}{\left(\frac{1}{2}\right)^{\frac{1}{2}}}\right)^{f_j(\frac{n}{2})}\nonumber\\
&=&\left(\sqrt{\frac{2}{e}}\right)^{f_j(\frac{n}{2})}\nonumber\\
&\leq&e^{-0.06f_j(\frac{n}{2})}\nonumber\\
&\leq&e^{-Cf_{min}(n)}\nonumber
\end{eqnarray}
where $C=\frac{0.06}{C_2}>0$, the last inequality is correct due to our assumption (1).\\
Combining (1) and (2) in our assumption \ref{assumption}, we know $\min_{1\leq j\leq M}{T_j}\geq\Omega\left(\max\{f_{min}(n)^{1-\alpha}, \log n\}\right)$ with at least the probability $1-Me^{-Cf_{min}(n)}$.
Therefore, our estimation error is bounded as\\
\begin{equation}
    e_\nu(n, \hat{\Delta}_n)\leq\mathcal{O}\left(\frac{1}{\max\{f_{min}(n)^{1-\alpha}, \log n\}}\right)+Me^{-Cf_{min}(n)}\mathcal{O}\left(1\right)=\mathcal{O}\left(\frac{1}{\max\{f_{min}(n)^{1-\alpha}, \log n\}}\right) \nonumber
\end{equation}
Thus, we finish the proof of theorem 2.\\

\subsection{Proof to Theorem \ref{theorem 3}}
The following proof is under the event $\bigcap_{1\leq j\leq M}(T_j\geq \frac{1}{2}T_{min})$, which probability is at least $1-\frac{M}{n^2}$.\\
Note that for any feature $X_j(1\leq j\leq M)$, we learn at least $T_j=\min\{f_j(n)-f_j(\frac{n}{2}), \frac{1}{2}T_{min}\}$ periods with equal probabilities of two arms, so the MSE estimation error of feature $X_j$ is bounded as $\mathcal{O}\left(\frac{1}{T_j}\right)$. Hence, our estimation error is bounded as $\mathcal{O}\left(\frac{1}{\min_{1\leq j\leq M}{T_j}}\right)$.\\
Now we focus on $T_j$.\\
For any $1\leq j\leq M$ and $1\leq t\leq n$, notice the characteristic function $I_{\{x_t=X_j\}}\in \{0,1\}$ and follows Bernoulli($p_j^t$), we have $\mathbb{E}\left[\hat{f_j}\right]=f_j(\frac{n}{2})=\sum_{1\leq t\leq \frac{n}{2}}p_j^t$\\
Therefore, by Chernoff bound (multiplicative form (relative error)), we have\\
\begin{eqnarray}
\mathbb{P}\left(\hat{f_j}<\frac{f_j(\frac{n}{2})}{2}\right)
&\leq& \left(\frac{e^{-\frac{1}{2}}}{\left(\frac{1}{2}\right)^{\frac{1}{2}}}\right)^{f_j(\frac{n}{2})}\nonumber\\
&=&\left(\sqrt{\frac{2}{e}}\right)^{f_j(\frac{n}{2})}\nonumber\\
&\leq&e^{-0.06f_j(\frac{n}{2})}\nonumber\\
&\leq&e^{-Cf_{min}(n)}\nonumber
\end{eqnarray}
where $C=\frac{0.06}{C_2}>0$, the last inequality is correct due to our assumption (1).\\
Combining (1) and (2) in our assumption \ref{assumption}, we know $\min_{1\leq j\leq M}{T_j}\geq\Omega\left(\max\{f_{min}(n)^{1-\alpha}, \frac{\log n}{\varepsilon}\}\right)$ with at least the probability $1-Me^{-Cf_{min}(n)}$.
Therefore, our estimation error is bounded as
$\mathcal{O}\left(\frac{1}{\max\{f_{min}(n)^{1-\alpha}, \frac{\log n}{\varepsilon}\}}\right)+Me^{-Cf_{min}(n)}\mathcal{O}\left(1\right)+\frac{M}{n^2}\mathcal{O}\left(1\right)=\mathcal{O}\left(\frac{1}{\max\{f_{min}(n)^{1-\alpha}, \frac{\log n}{\varepsilon}\}}\right)$.
\subsection{Proof to Lemma \ref{lemma 1}, Corollary \ref{corollary 1},  Lemma \ref{lemma 2} and Theorem \ref{theorem-dpupper}}
\subsubsection{Proof of lemma \ref{lemma 1}}
\ \\
Given an epoch $e$ we denote by $\mathcal{E}_e$ the event where for all arms $a \in S$ it holds that (we use $T:=n_j$ represents the number of occurrences of the feature $X_j$ and $\beta=\frac{1}{n}$ below)\\
(i) $\left|\mu_a-\bar{\mu}_a\right| \leq h_e$;\\
(ii) $\left|\bar{\mu}_a-\widetilde{\mu}_a\right| \leq c_e$;\\
(iii) $R_e\leq R^j_e\leq 3R_e$;\\
and also denote $\mathcal{E}=\bigcap_{e \geq 1} \mathcal{E}_e$.\\
First, by definition, we can calculate that:\\
$R_1\geq \frac{16\log T}{\epsilon}$, so $R_e\geq 2R_{e-1}\geq \frac{2^{e+3}\log T}{\epsilon}$.\\
Hence, $P((iii)^c)\leq 2exp\{-R_e\epsilon\}\leq 2T^{-2^{e+3}}$\\
Furthermore, given (iii), the Hoeffding bound, concentration of the Laplace distribution and the union bound over all arms in $S_0$ give that $\operatorname{Pr}\left[\mathcal{E}_e\right] \geq 1-\left(\frac{\beta}{4 e^2}+\frac{\beta}{4 e^2}+2T^{-2^{e+3}}\right)$, thus $\operatorname{Pr}[\mathcal{E}] \geq 1-\frac{\beta}{2}\left(\sum_{e \geq 1} e^{-2}\right)-\sum_{e \geq 1}2T^{-2^{e+3}} \geq 1-\frac{1}{T}$ ($T\geq 3$). The remainder of the proof continues under the assumption the $\mathcal{E}$ holds, and so, for any epoch $e$ and any viable arm $a$ in this epoch we have $\left|\widetilde{\mu}_a-\mu_a\right| \leq h_e+c_e$. As a result for any epoch $e$ and any two arms $a^1, a^2$ we have that $\left|\left(\widetilde{\mu}_{a^1}-\widetilde{\mu}_{a^2}\right)-\left(\mu_{a^1}-\mu_{a^2}\right)\right| \leq 2 h_e+2 c_e$.

Next, we argue that under $\mathcal{E}$ the optimal arm $a^*$ is never eliminated. Indeed, for any epoch $e$, we denote the arm $a_e=\operatorname{argmax}_{a \in S} \widetilde{\mu}_a$ and it is simple enough to see that $\widetilde{\mu}_{a_e}-\widetilde{\mu}_{a^*} \leq$ $0+2 h_e+2 c_e$, so the algorithm doesn't eliminate $a^*$.

Next, we argue that, under $\mathcal{E}$, in any epoch $e$ we eliminate all viable arms with suboptimality gap $\geq 2^{-e}=\Delta_e$. Fix an epoch $e$ and a viable arm $a$ with suboptimality gap $\Delta_a \geq \Delta_e$. Note that we have set parameter $R_e$ so that
$$
\begin{aligned}
h_e & =\sqrt{\frac{\log \left(16 \cdot e^2 / \beta\right)}{2 R_e}}<\sqrt{\frac{\log \left(16 \cdot e^2 / \beta\right)}{2 \cdot \frac{32 \log \left(16 e^2 / \beta\right)}{\Delta_e^2}}}=\frac{\Delta_e}{8};\\
c_e & =\frac{\log \left(8 \cdot e^2 / \beta\right)}{R_e \varepsilon}<\frac{\log \left(8 \cdot e^2 / \beta\right)}{\varepsilon \cdot \frac{8 \log \left(8 e^2 / \beta\right)}{\varepsilon \Delta_e}}=\frac{\Delta_e}{8}
\end{aligned}
$$

Therefore, since arm $a^*$ remains viable, we have that $\widetilde{\mu}_{\max }-\widetilde{\mu}_a \geq \widetilde{\mu}_{a^*}-\widetilde{\mu}_a \geq \Delta_a-\left(2 h_e+2 c_e\right)>$ $\Delta_e\left(1-\frac{2}{8}-\frac{2}{8}\right) \geq \frac{\Delta_e}{2}>2 h_e+2 c_e$, guaranteeing that arm $a$ is removed from $S$.

Lastly, fix a suboptimal arm $a$ and let $e(a)$ be the first epoch such that $\Delta_a \geq \Delta_{e(a)}$, implying $\Delta_{e(a)} \leq \Delta_a<\Delta_{e(a)-1}=2 \Delta_e$. Using the immediate observation that for any epoch $e$ we have $R_e \leq R_{e+1} / 2$, we have that the total number of pulls of arm $a$ is
$$
\sum_{e \leq e(a)} R^j_e \leq 3\sum_{e \leq e(a)} R_e \leq 3\sum_{e \leq e(a)} 2^{e-e(a)} R_{e(a)} \leq 3R_{e(a)} \sum_{i \geq 0} 2^{-i} \leq 6\left(\frac{32 \log \left(16 \cdot e(a)^2 / \beta\right)}{\Delta_e^2}+\frac{8 \log \left(8 \cdot e(a)^2 / \beta\right)}{\varepsilon \Delta_e}\right)
$$

The bounds $\Delta_e>\Delta_a / 2,|S| \leq 2, e(a)<\log _2\left(2 / \Delta_a\right)$ allow us to conclude and infer that under $\mathcal{E}$ the total number of pulls of arm $a$ is at most
\begin{equation}
    3\log \left(2 \log \left(2 / \Delta_a\right) / \beta\right)\left(\frac{1024}{\Delta_a^2}+\frac{96}{\varepsilon \Delta_a}\right)=\mathcal{O}\left(\left(\log n_j+\log \log \left(1 / \Delta(X_j)\right)\right)\left(\frac{1}{\Delta(X_j)^2}+\frac{1}{\varepsilon \Delta(X_j)}\right)\right) \nonumber
\end{equation}
\subsubsection{Proof of corollary \ref{corollary 1}}
\ \\
By using the result of lemma \ref{lemma 1}, for $n>\log 1/\Delta(X_j)$, we have
\begin{equation*}
\begin{aligned}
    \mathcal{R}^{first}_\nu(n, \pi)&\leq \Sigma_{1\leq j\leq M}\Delta(X_j)\mathcal{O}\left(\left(\log n_j+\log \log \left(1 / \Delta(X_j)\right)\right)\left(\frac{1}{\Delta(X_j)^2}+\frac{1}{\varepsilon \Delta(X_j)}\right)\right)\\
    &\leq \mathcal{O}\left(\left(\sum_{1 \leq j\leq M} \frac{\log n}{\Delta(X_j)}\right)+\frac{M\log n}{\varepsilon}\right)
\end{aligned}
\end{equation*}

\subsubsection{Proof of lemma \ref{lemma 2}}
\ \\
Noticed that with the probability $\geq 1-\frac{1}{n}$, the optimal arm would be chosen correctly in the first half periods. Therefore, the expected regret of the second half periods of algorithm 2 is\\
$\mathcal{R}^{second}_\nu(n, \pi)\leq \Sigma_{1\leq j\leq M}\Delta(X_j)\mathbb{E}\left[T_j\right]+\frac{1}{n}\mathcal{O}\left(n\right)=\mathcal{O}\left(\max\{f_{min}(n)^{1-\alpha}, \frac{\log n}{\varepsilon} \}\sum_{1 \leq j\leq M} \Delta(X_j)\right)$\\

\subsubsection{Proof of theorem \ref{theorem-dpupper}}
\ \\
By adding the result of corollary 1 and lemma 2, under the condition that $\Delta(X_j)=\mathcal{O}(1)$ $(\forall 1\leq j\leq M)$ we can easily proof that $\mathcal{R}_\nu(n, \pi)\leq \mathcal{O}\left(M\max\{f_{min}(n)^{1-\alpha}, \frac{\log n}{\varepsilon} \}\right)$.\\
As a result, we have $e_\nu(n, \hat{\Delta}_n) \mathcal{R}_\nu(n, \pi)\le \mathcal{O}(M)$, which is the same as theorem \ref{theorem-upper}, and matches the lower bound in theorem \ref{thm-lower}.\\
Finally, we need to prove that the algorithm 2 is $\left(\varepsilon, \frac{1}{n}\right)$-private.\\
The following proof is under the event $\bigcup_{1\leq j\leq M}\bigcup_{e \geq 1}(R^j_e\geq R_e)$, which probability is at least $1-\frac{1}{n}$ $(n\geq3M)$.\\
For any two neighboring datasets D and D', suppose D and D' are only different at time $t$.We discuss different cases for $t$ as following:\\
(1) $t$ is in the second half, i.e. $[\frac{n}{2}]+1\leq t\leq n$;\\
In this case, since the probabilities of arms(actions) are not dependent of features or rewards (always $(\frac{1}{2}, \frac{1}{2})$, and noticed that the output $\Delta$ and running time periods $T_j$ are added Laplace mechanism, which are both $\frac{\epsilon}{2}$-private. Therefore, in this case, $P(D)\leq e^\epsilon P(D')$.
(2)$t$ is in the first half, i.e. $1\leq t\leq [\frac{n}{2}]$;\\
Noticed that $|T_{min}-T'_{min}|\leq 1$ and $T_j$ and $T'_j$ are added  Laplace mechanism, which is $\frac{\epsilon}{2}$-private for the second half.\\
Moreover, for the first half, the mean values $\mu$s and each running periods are all added Laplace mechanism, which are  $\frac{\epsilon}{2}$-private.\\
Therefore, by the composition theorem, $P(D)\leq e^\epsilon P(D')$.\\
In conclusion, for any two neighboring datasets D and D', we have $P(D)\leq e^\epsilon P(D')+\frac{M}{n^2}\leq e^\epsilon P(D')+\frac{1}{n}$
\subsection{Proof of Proposition \ref{CLT 1} and Proposition \ref{CLT 2}}
\subsubsection{Proof of proposition \ref{CLT 1}}
\ \\
By our definition and Central Limit Theorem (CLT), we know $T_{min}=\max\{f_{min}(n)^{1-\alpha}, \log n \}$ and $\lim_{n \to \infty} \sqrt{T_{min}}(\hat{\Delta}(X_j)-\Delta(X_j)) \leadsto \textbf{N}(0, \sigma_{j0}^2+\sigma_{j1}^2)$.
\subsubsection{Proof of proposition \ref{CLT 2}}
\ \\
We have $T_{min}=\max\{f_{min}(n)^{1-\alpha}, \frac{\log n}{\varepsilon} \}$ and $\sqrt{\max\{f_{min}(n)^{1-\alpha}, \frac{\log n}{\varepsilon} \}}(\hat{\Delta}(X_j)-\Delta(X_j))=\sqrt{T_{min}}\left((\bar{\mu}_1^j-\bar{\mu}_0^j)-(\mu_1^j-\mu_0^j)+Lap(2/\varepsilon T_j)\right)$.\\
To prove the above expression is asymptotically normal, we will give the proof of the three solutions below:\\
(1) $\frac{T_j}{T_{min}}\xrightarrow{P}1$.\\
(2) $\sqrt{T_{min}}Lap(2/\varepsilon T_j)\xrightarrow{P}0$.\\
(3) $\sqrt{T_{min}}\left((\bar{\mu}_1^j-\bar{\mu}_0^j)-(\mu_1^j-\mu_0^j)\right)\xrightarrow{L}\textbf{N}(0, \sigma_{j0}^2+\sigma_{j1}^2 )$.\\
For the solution (1), by our definition of $T_j$, we know for any $\delta>0$,
$$
P\left(\left|\frac{T_j}{T_{min}}-1\right| \geq \delta \right)\leq 2\sum_{k\geq \delta T_{min}}e^{-\frac{\varepsilon}{2}|k|}\rightarrow0
$$
as $T_{min}\geq \log n \rightarrow +\infty$.\\
For the solution (2), by using the solution (1), we know for any $\delta\in (0,1)$,
$$
P\left(|X|>\delta\right)\leq P\left(\left|\frac{T_j}{T_{min}}-1\right| \geq \delta \right)+P\left(|X|>\delta, \left|\frac{T_j}{T_{min}}-1\right| \leq \delta \right)
\leq P\left(\left|\frac{T_j}{T_{min}}-1\right| \geq \delta \right)+ e^{-(1-\delta)\delta\sqrt{T_{min}}\epsilon}
\rightarrow 0
$$
as $T_{min}\geq \log n \rightarrow +\infty$, where r.v. $X$ follows the distribution $\sqrt{T_{min}}Lap(2/\varepsilon T_j)$.\\
For the solution (3), denote $S_{T_j}=T_j\left((\bar{\mu}_1^j-\bar{\mu}_0^j)-(\mu_1^j-\mu_0^j)\right)$ is the sum of $T_j$ i.i.d differences. Similarly, we denote $S_{T_{min}}$ is the sum of $T_{min}$ i.i.d. differences who follow the same distribution with expectation 0 and variance $\sigma_{j0}^2+\sigma_{j1}^2$.\\
Obviously, by Central Limit Theorem (CLT), we have $\frac{S_{T_{min}}}{\sqrt{T_{min}}}\xrightarrow{L}\textbf{N}(0, \sigma_{j0}^2+\sigma_{j1}^2)$. Therefore, we only need to prove that
$$
\frac{S_{T_{min}}-S_{T_j}}{\sqrt{T_{min}}}\xrightarrow{P}0
$$
By using the solution (1), we know for any $\delta, \delta'>0$,
\begin{eqnarray}
  P\left(\left|\frac{S_{T_{min}}-S_{T_j}}{\sqrt{T_{min}}}\right|>\delta'\right) 
  &\leq& P\left(\left|\frac{T_j}{T_{min}}-1\right| \geq \delta \right)+P\left(\left|\frac{S_{T_{min}}-S_{T_j}}{\sqrt{T_{min}}}\right|>\delta', \left|\frac{T_j}{T_{min}}-1\right| \leq \delta \right) \nonumber \\
  &\leq& P\left(\left|\frac{T_j}{T_{min}}-1\right| \geq \delta \right) + \frac{\delta}{\delta'^2} (\sigma_{j0}^2+\sigma_{j1}^2) \nonumber \\
  &\rightarrow& 0 \nonumber
\end{eqnarray}
as $T_{min}\geq \log n \rightarrow +\infty$ and letting $\delta\rightarrow0$, where the last inequality is because of Markov inequality. \\
Combining solutions (2) and (3), we can easily know the proposition 2 is correct.
\end{document}